\documentclass[twocolumn]{aastex631}

\usepackage{gensymb}
\usepackage{amsmath}
\usepackage{comment}

\shorttitle{Gaia24ccy: An outburst followed the footsteps of its predecessor}
\shortauthors{Singh et al.}

\graphicspath{{./}{figures/}}

\begin{document}
\title{Gaia24ccy: An outburst followed the footsteps of its predecessor}

\correspondingauthor{Koshvendra Singh}
\email{koshvendra1999@gmail.com}
\author[0000-0002-7434-9681]{Koshvendra Singh}
\affiliation{Department of Astronomy and Astrophysics, Tata Institute of Fundamental Research, Mumbai, 400005, India}

\author[0000-0001-8720-5612]{Joe P. Ninan}
\affiliation{Department of Astronomy and Astrophysics, Tata Institute of Fundamental Research, Mumbai, 400005, India}

\author[0000-0003-0292-4832]{Zhen Guo}
\affiliation{Instituto de F\'isica y Astronom\'ia, Universidad de Valpara\'iso, Av. Gran Breta\~na 1111, Playa Ancha, Casilla 5030, Chile}
\affiliation{Millennium Institute of Astrophysics, Nuncio Monse{\~n}or Sotero Sanz 100, Of. 104, Providencia, Santiago, Chile}

\author[0000-0002-5963-1283]{Valentin D. Ivanov}
\affiliation{European Southern Observatory, Karl-Schwarzschild-Strasse 2, D-85748 Garching bei M\"unchen, Germany}

\author[0000-0002-7004-9956]{David A. H. Buckley}
\affiliation{South African Astronomical Observatory, PO Box 9, Observatory 7935, Cape Town, South Africa}
\affiliation{Department of Astronomy, University of Cape Town, Private Bag X3, Rondebosch 7701, South Africa}
\affiliation{Department of Physics, University of the Free State, PO Box 339, Bloemfontein 9300, South Africa}

\author[0000-0001-9312-3816]{Devendra K. Ojha}
\affiliation{Department of Astronomy and Astrophysics, Tata Institute of Fundamental Research, Mumbai, 400005, India}

\author[0000-0002-0048-2586]{Andrew Monson}
\affiliation{Steward Observatory and Department of Astronomy, University of Arizona, 933 N. Cherry Avenue, Tucson, AZ 85721, USA}

\author[0009-0008-8490-8601]{Tarak Chand}
\affiliation{Aryabhatta Research Institute of Observational Sciences (ARIES), Manora Peak, Nainital-263001, India}
\affiliation{M.J.P. Rohilkhand University, Bareilly-243006, India}

\author[0000-0001-5731-3057]{Saurabh Sharma}
\affiliation{Aryabhatta Research Institute of Observational Sciences (ARIES), Manora Peak, Nainital-263001, India}

\author[0000-0002-6740-7425]{Ram Kesh Yadav}
\affiliation{National Astronomical Research Institute of Thailand (public organization), 260 Moo 4, T. Donkaew, A. Maerim, Chiangmai, 50180, Thailand}

\author[0000-0002-6688-0800]{Devendra K. Sahu}
\affiliation{Indian Institute of Astrophysics, II Block, Koramangala, Bangalore 560034, India}

\author{Pramod Kumar}
\affiliation{Indian Institute of Astrophysics, II Block, Koramangala, Bangalore 560034, India}

\author[0000-0002-9433-900X]{Vardan Elbakyan}
\affiliation{Fakult\"at f\"ur Physik, Universit\"at Duisburg-Essen, Lotharstraße 1, D-47057 Duisburg, Germany}
\affiliation{Research Institute of Physics, Southern Federal University, Rostov-on-Don 344090, Russia}

\author[0000-0002-6166-2206]{Sergei Nayakshin}
\affiliation{School of Physics, University of Leicester, Leicester, LE1 7RH, UK}

\author[0009-0003-4432-9537]{Vitor Fermiano}
\affiliation{Departamento de F\'isica, Universidade Federal de Santa Catarina, Trindade 88040-900 Florian\'opolis, Brazil}

\author[0000-0001-8060-1321]{Min Fang}
\affiliation{Purple Mountain Observatory, Chinese Academy of Sciences, 10 Yuanhua Road, Nanjing 210023, People's Republic of China}
\affiliation{School of Astronomy and Space Science, University of Science and Technology of China, 96 Jinzhai Road, Hefei 230026, People's Republic of China}

\author[0000-0002-5936-7718]{Jura Borissova}
\affiliation{Instituto de F\'isica y Astronom\'ia, Universidad de Valpara\'iso, Av. Gran Breta\~na 1111, Playa Ancha, Casilla 5030, Chile}
\affiliation{Millennium Institute of Astrophysics, Nuncio Monse{\~n}or Sotero Sanz 100, Of. 104, Providencia, Santiago, Chile}

\author[0000-0003-0262-272X]{Wen Ping Chen}
\affiliation{Institute of Astronomy, National Central University, 300 Zhongda Road, Zhongli 32001 Taoyuan, Taiwan}

\author[0000-0003-0125-8700]{Franz-Josef Hambsch}
\affiliation{Vereniging Voor Sterrenkunde (VVS), Zeeweg 96, 8200 Brugge, Belgium}
\affiliation{Groupe Europ\'een d'Observations Stellaires (GEOS), 23 Parc de Levesville, 28300 Bailleau l\`Ev\^eque, France }
\affiliation{Bundesdeutsche Arbeitsgemeinschaft f\"ur Ver\"anderliche Sterne (BAV), Munsterdamm 90, 12169 Berlin, Germany}

\author[0000-0002-9740-9974]{Radostin Kurtev}
\affiliation{Instituto de F\'isica y Astronom\'ia, Universidad de Valpara\'iso, Av. Gran Breta\~na 1111, Playa Ancha, Casilla 5030, Chile}
\affiliation{Millennium Institute of Astrophysics, Nuncio Monse{\~n}or Sotero Sanz 100, Of. 104, Providencia, Santiago, Chile}

\author[0000-0002-7535-7077]{Calum Morris}
\affiliation{Instituto de F\'isica y Astronom\'ia, Universidad de Valpara\'iso, Av. Gran Breta\~na 1111, Playa Ancha, Casilla 5030, Chile}
\affiliation{Millennium Institute of Astrophysics, Nuncio Monse{\~n}or Sotero Sanz 100, Of. 104, Providencia, Santiago, Chile}

\author[0009-0007-6769-7075]{Javier Osses}
\affiliation{Instituto de F\'isica y Astronom\'ia, Universidad de Valpara\'iso, Av. Gran Breta\~na 1111, Playa Ancha, Casilla 5030, Chile}
\affiliation{Millennium Institute of Astrophysics, Nuncio Monse{\~n}or Sotero Sanz 100, Of. 104, Providencia, Santiago, Chile}

\author[0009-0007-6273-9141]{Vania Rodr\'iguez}
\affiliation{Instituto de F\'isica y Astronom\'ia, Universidad de Valpara\'iso, Av. Gran Breta\~na 1111, Playa Ancha, Casilla 5030, Chile}

\author[0000-0003-1634-3158]{Tanvi Sharma}
\affiliation{Institute of Astronomy, National Central University, 300 Zhongda Road, Zhongli 32001 Taoyuan, Taiwan}

\author[0009-0004-1054-2812]{Bandari Srikanth}
\affiliation{Indian Institute of Astrophysics, II Block, Koramangala, Bangalore 560034, India}

\author[0000-0003-4507-1710]{Thanawuth Thanathibodee}
\affiliation{Department of Physics, Faculty of Science, Chulalongkorn University, 254 Phayathai Road, Pathumwan, Bangkok 10330, Thailand}

\author[0000-0003-2588-1265]{Wei-Hao Wang}
\affiliation{Institute of Astronomy and Astrophysics, Academia Sinica, Taipei 10617, Taiwan}

\author[0009-0004-8171-3551]{Yuting Zhou}
\affiliation{High School Affiliated to Nanjing Normal University, 210024, Nanjing, Jiangsu Province, China}

\begin{abstract}

Accretion-driven outbursts in young stellar objects remain poorly understood, largely limited by a statistically small sample of closely followed-up events. This underscores the importance of a thorough exploration of each outbursting object. We studied a peculiar outbursting system, Gaia24ccy, which exhibited two $\Delta g\sim3.8$ mag outbursts in 2019 and 2024. The system consists of two unresolved, nearly identical, and rapidly rotating young stars: Gaia24ccy\,A (1.1419 days) and Gaia24ccy\,B (1.7898 days). Periodogram analyses just before the onset of the outbursts suggest Gaia24ccy\,B to be the outbursting component. Unlike any previously known EXor sources, the two outburst profiles show a very similar evolution: both rose at the same rate for the first 15 days, followed by many `\textit{sub-bursts}' on the timescale of 10$-$20 days. The 2019 outburst lasted 145$-$255 days, while the 2024 outburst persisted for 367 days. 
We infer the unstable region to lie at $r_{\rm trigger}\simeq0.019-0.047$ au ($\sim5-12.3 R_\star$). The accreted mass per event $M_{\rm acc}\sim10^{-5} M_\odot$ can be provided by a compact inner–disk reservoir. The photometric rise/decay timescales and the mid-infrared color evolution favor a thermal–viscous trigger in a hot inner disk, while the appearance of rich emission-line spectra indicates concurrent magnetospheric compression — together best described by a hybrid picture. Finally, we explain the reddening of the mid-infrared color observed during the outburst as a consequence of the competing emission from the viscous disk and the photosphere.

\end{abstract}

\keywords{accretion, accretion-disks — stars: low-mass — stars: individual (2MASS J16133650-2503473)}

\section{Introduction} \label{sec:intro}

Accretion-driven luminosity outbursts in the young stellar objects (YSOs) have been proposed as solutions to various longstanding problems in low-mass star formation. These include the presence of crystalline silicates in our solar system's cometary objects \citep{2009Natur.459..224A}, the luminosity problem $-$ where the observed accretion rates are lower than expected from disk evolution timescales \citep{1990AJ.....99..869K}, the luminosity spread among YSOs \citep{2023ASPC..534..355F}, and the presence of knots in the YSO outflows \citep{2012MNRAS.425.1380I}. The outbursts also provide a unique window to the innermost dynamics of YSOs \citep{2024ApJ...968...88S}.

However, what triggers these accretion bursts remains poorly understood. Various existing hypotheses explain and predict certain timescales and flux scales based on the triggering mechanisms, which can be confronted with observations. One of the initial hypotheses was thermal instability developed by the runaway viscous heat-trapping by Hydrogen ionization in the accretion disk, leading to self-regulated accretion-bursts \citep{1994ApJ...427..987B,2024MNRAS.530.1749N}. \cite{1992ApJ...401L..31B,2016ApJ...827...43M} showed that circumstellar binaries in an eccentric orbit around each other would produce accretion bursts at periastron \citep{2017ApJ...842L..12T}. DQ Tau, a close binary system, has shown similar bursts \citep{2018ApJ...862...44K}. Similarly, massive planets eccentrically orbiting the central accreting YSO, at its inner disk, can produce accretion bursts once or twice during the rotation period \citep{2018ApJ...853L..34B,2020MNRAS.495.3920T}. Tidal disruption of a planet at the very inner edge of the disk could lead to a sustained accretion outburst \citep{2012MNRAS.426...70N,2024MNRAS.528.2182N}. Stellar flybys can also trigger extreme accretion events \citep{2023EPJP..138...11C}. Another possibility presented by \citet{2012MNRAS.420..416D} is trapping the disk material at the outer edge of the magnetosphere, which builds up until ram pressure exceeds the magnetospheric pressure. Change in the stellar dynamo cycle has also been suggested as a possible cause of the accretion outburst \citep{2016ApJ...833L..15A}.

Many of these hypotheses predict periodic outbursts, highlighting such YSOs as key cases of interest. \citet{2022MNRAS.513.1015G} found 59 periodically outbursting YSOs in the near-infrared (NIR) photometric survey from the 2nd version of VISTA Variables in the Via Lactea (VVV) Infrared Astrometric Catalogue \citep[VIRAC2-$\beta$,][]{2018MNRAS.474.1826S,2025MNRAS.536.3707S}.

\begin{figure*}
    \centering
    \includegraphics[width=1\textwidth]{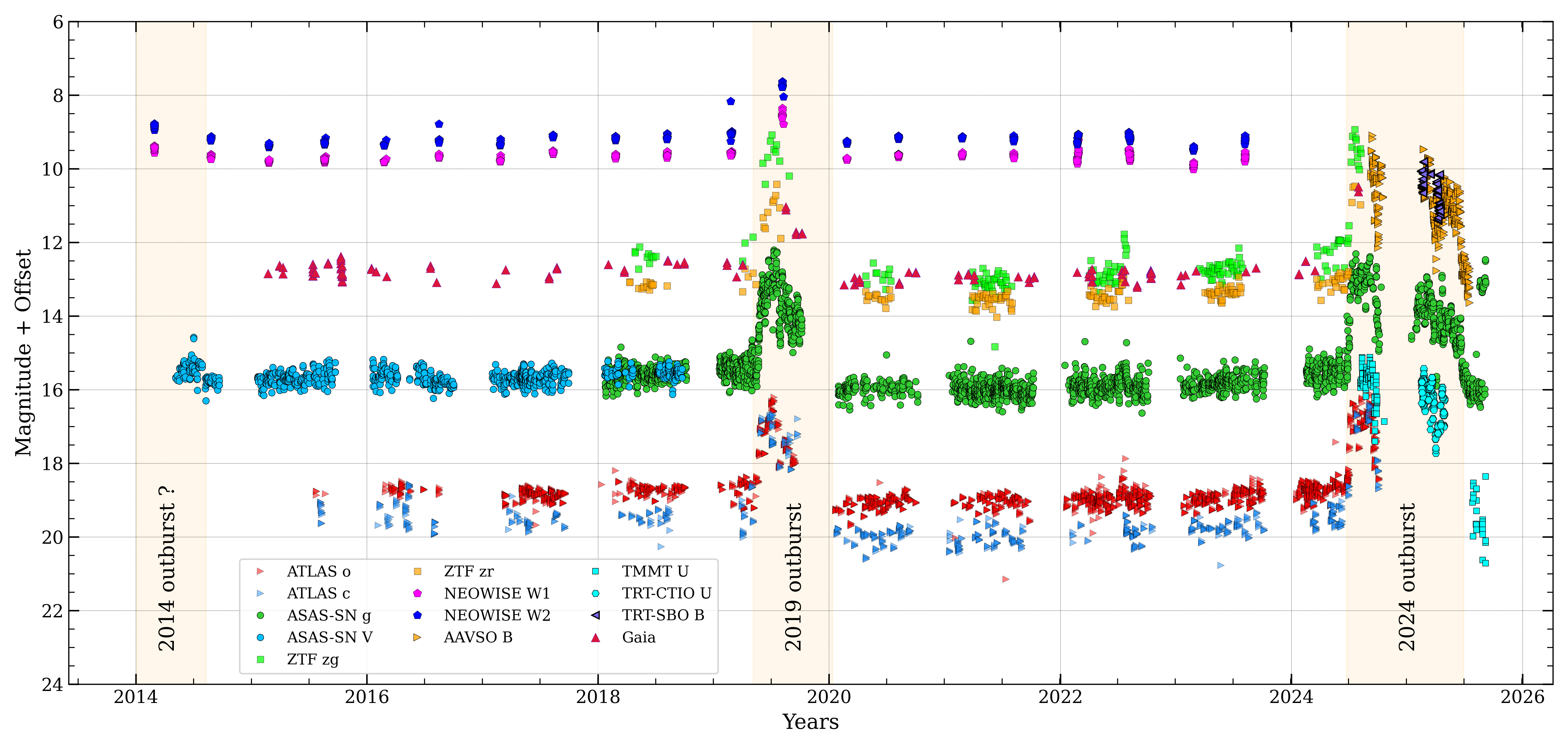}
    \includegraphics[width=1\textwidth]{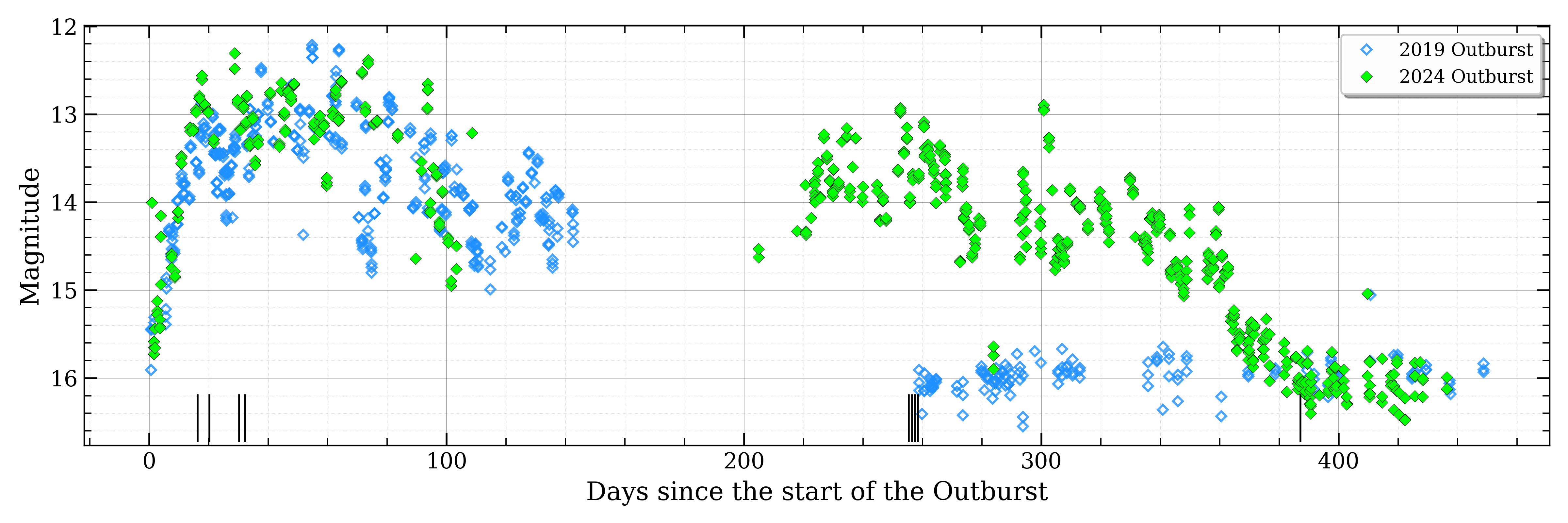}
    \caption{\textit{Upper panel}: Multiband light curves of Gaia24ccy, shown with magnitude offsets applied for clarity of multiwavelength evolution. Offsets are: ATLAS $o$ (+4.5), ATLAS $c$ (+4), ASAS-SN $V$ (+0.5), ZTF $zr$ (-1.5), ZTF $zg$ (-3.5), AAVSO $B$ (-4), TMMT $U$ (+2.5), TRT-CTIO $U$ (+2.5), TRT-SBO $B$ (-4). The durations of the 2019 and 2024 outbursts are labeled and shaded with a light orange color. Other band light curves of AAVSO, TMMT, TRT-CTIO, and TRT-SBO are not shown for the clarity of presentation. \textit{Lower panel}: The evolution of 2019 and 2024 outbursts is shown with ASAS-SN $g$ light curves. The start dates for the 2019 and 2024 outbursts are JD = 2458620, and 2460485, respectively. Grey vertical lines at the bottom mark the epochs of HFOSC observations during the 2024 outburst. The photometry from this work, TMMT, LCOGT, and TRT, is available in machine readable format as the data behind the figure.}
    \label{fig:lightcurve}
\end{figure*}

2MASS J16133650-2503473 is one such object, a promising candidate for exhibiting periodic outbursts. On 2024 July 08, it was reported in the Astronomers' Telegram \citep[ATel \#16696][]{2024ATel16696....1S} as a photometrically rising source in ASAS-SN $g$-band. The Gaia photometric alert system also flagged this event with the identifier Gaia24ccy. The ALeRCE broker \citep[Automatic Learning for the Rapid Classification of Events,][]{2021AJ....161..141S,2021AJ....161..242F} identified this rising object as a YSO. 

Gaia24ccy subsequently exhibited an outburst with $\Delta g \approx$ 3.8 mag. We also report here for the first time a previously overlooked event in 2019 with essentially the same amplitude. Furthermore, the object showed a rising mid-infrared flux (MIR, $W1$ and $W2$) around 2014, accompanied by a $\Delta V\sim1$ mag bump (see Figure \ref{fig:lightcurve}). Figure \ref{fig:lightcurve} highlights the similarity of the 2019 and 2024 outbursts. Gaia24ccy underwent two very similar outbursts after $\sim$5 years, with another possible candidate in 2014 (if missed), at an interval of 5 years. Thus, this object is an excellent candidate to investigate the mechanisms triggering periodic accretion outbursts.

\begin{figure}
    \centering
    \includegraphics[scale=0.4]{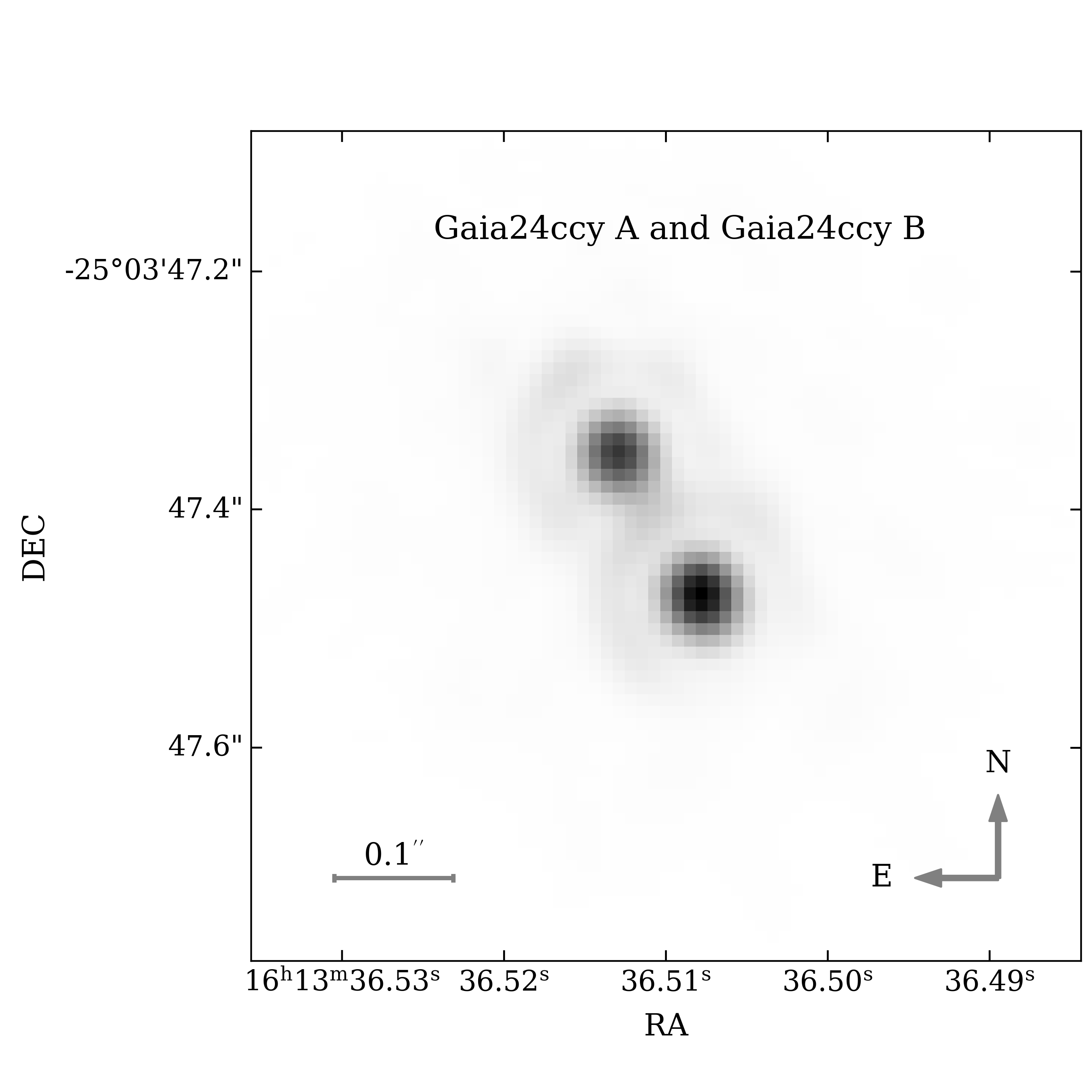}
    \caption{The $K$-band AO image of Gaia24ccy with Near-infrared Camera, Generation II (NIRC2) onboard Keck. The image is reproduced with permission from the authors \citep{2019ApJ...878...45B}. The two objects with rotation periods of 1.1419 and 1.7898 days are named Gaia24ccy\,A and Gaia24ccy\,B, respectively. Without spatially resolved rotation data, the A/B identification of the upper versus lower component remains undetermined.}
    \label{fig:KeckAO}
\end{figure}

Gaia24ccy resides in the Upper Scorpius OB association, a $\sim$8-10 Myr old star-forming region \citep{2012ApJ...746..154P,2016A&A...593A..99F,2019ApJ...872..161D}, at coordinates ($\alpha_{\mathrm{J2000}}:  16^{\mathrm hr}13^{\mathrm m}36.507^{\mathrm s}$, $\delta_{\mathrm{J2000}}: -25^{\mathrm d}03^{\mathrm m}47.736^{\mathrm s}$). Since Gaia Early Data Release 3 \citep[EDR3;][]{2021AJ....161..147B} did not provide parallax measurements to the source, \citet{2023ApJ...945..112F} adopted the mean distance to the Upper Scorpius region, 142 pc. We adopt the same distance in this study. \textit{K}-band Adaptive Optics (AO) imaging of this source revealed that Gaia24ccy consists of two point-like objects at a separation of 138.4 $\pm$ 1.8 milliarcseconds (mas) \citep[$\equiv$ 19.7 au;][]{2019ApJ...878...45B}, see Figure \ref{fig:KeckAO}. The two objects are of similar near infrared (NIR) brightness, $\Delta K$=0.26 mag. \citet{2017ApJ...851...85B} modeled the 0.88 mm dust continuum emission of Gaia24ccy from ALMA observation and derived a disk inclination angle of 86$^{\circ +4}_{-52}$. Considering it as a single object, the authors also estimated the outer disk radius of Gaia24ccy to be 45$^{-48}_{-33}$ au. The disk could be nearly edge-on, as also concluded by \citet{2024A&A...688A..61N} in the study of dippers as a consequence of dust rising along the magnetospheric funnel. \citet{2023ApJ...945..112F} estimated the extinction towards the object to be $A_{V}$=1.3 mag.

The paper is structured as follows: Section \ref{sec:data reduc} discusses the spectroscopic and photometric data used in the work and elaborates on the data reduction and calibration processes. Results from the data are presented and analyzed in Section \ref{sec:analysis}. We discuss and try to converge on a picture of the star-disk system and outburst phenomenon in Section \ref{sec:discussion}. We conclude this work by highlighting key findings in Section \ref{sec:conclusion}. Appendices at the end of the paper provide further details on the analysis.

\section{Observations and Data Reduction} \label{sec:data reduc}

In this section, we discuss the spectroscopic and photometric data taken with various ground-based telescopes.

\subsection{HFOSC onboard 2m HCT}\label{subsec:data_reduc hfosc}
India hosts a 2m Himalayan Chandra Telescope (HCT) at Hanle, Ladakh, in the foothills of the Himalayas. Optical spectra were obtained with the Hanle Faint Object Spectrograph and Camera (HFOSC) on HCT. Spectra were obtained with Grating-8 (Gr8) and Grating-14 (Gr14) with a slit width of 167 $\mu$m, yielding a spectral resolution of 1200 and 1320, respectively. Gr8 and Gr14 have a wavelength coverage of 5800-9200 \r{A} and 3270-6160 \r{A}, respectively. The log of observation along with signal-to-noise ratio (SNR) is provided in Table \ref{table:obs_log}. The single-order spectra were reduced with the Image Reduction and Analysis Facility \citep[IRAF;][]{1986SPIE..627..733T}. Wavelength calibration was performed using a FeNe lamp with a corresponding grating and 67 $\mu$m slit. The wavelength solution is further corrected by a linear fit to the Gaussian centers of various OI atmospheric lines (5577 \r{A}, 6300 \r{A}, 7913 \r{A}, 8344 \r{A}, and 8827 \r{A}).

We constructed an instrument response function (IRF) from HFOSC observation of the telluric standard stars Feige110 and Feige34, and utilizing their flux-calibrated spectra from \textit{CALSPEC} database. This IRF was used to correct all the spectra.

\begin{figure*}
    \centering
    \includegraphics[width=1\linewidth]{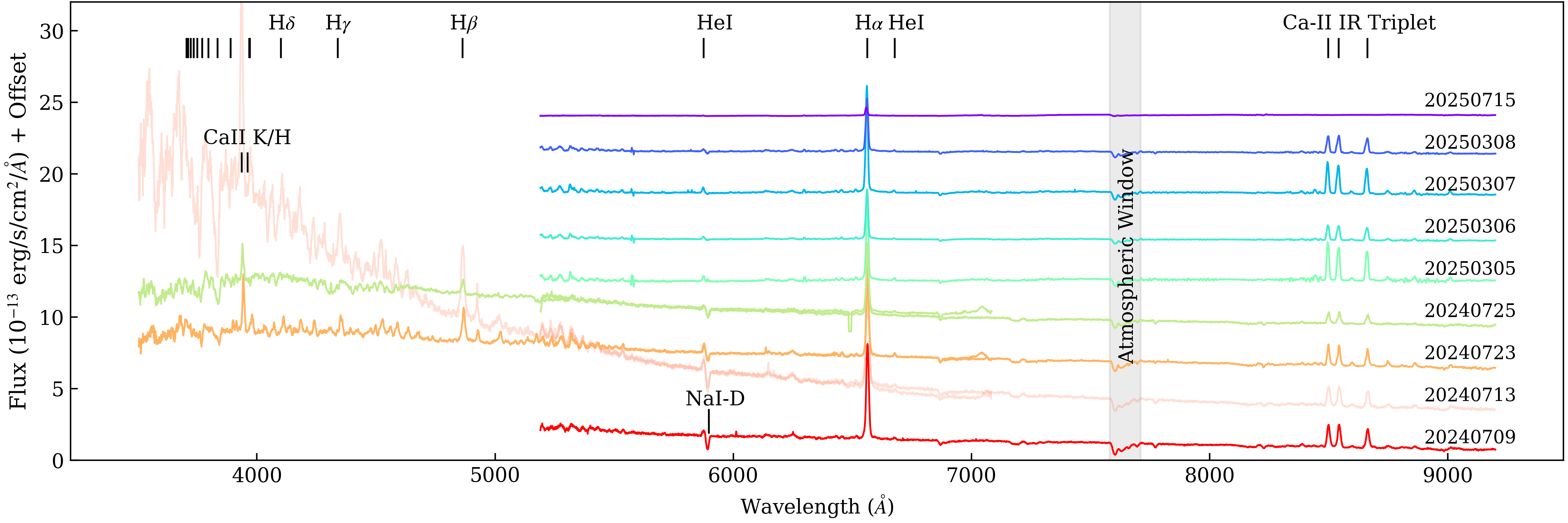}
    \caption{The flux-calibrated optical HFOSC spectra of Gaia24ccy are shown. A few prominent spectral lines are marked for reference.}
    \label{fig:HFOSCspectra}
\end{figure*}

We flux-calibrated the spectra by interpolating the ASAS-SN $g$-band photometric fluxes to the epochs of the spectra. We also analyzed other light curves to assess the photometric variability between the epochs of ASAS-SN $g$ observations. The 2024 July 13 spectrum shows an abnormally large flux towards shorter wavelengths (see Figure \ref{fig:HFOSCspectra}) and so it is excluded from all the analysis. The flux-calibrated spectra are shown in Figure \ref{fig:HFOSCspectra}.

\subsection{Three hundred MilliMeter Telescope (TMMT)}\label{subsec:TMMT}

We monitored Gaia24ccy during the first half of its outburst with TMMT \citep{2017AJ....153...96M} at Las Campanas Observatory, Chile. Photometric observations were performed at nearly one-day cadence in $U, B, V, R$ and $I$-bands. A total of 147, 163, 160, 159, and 162 frames were collected in respective bands during our observation campaign from 2024 August 06 to 2025 September 06. The aperture photometry of the frames was calibrated with the color equations developed from the observation of four Landolt fields: SA092, SA108, SA111, and SA113 \citep{1992AJ....104..340L,2022A&A...664A.109P}. \\

\begin{align*}
U - B &= 0.86(u - b) - 0.12 (X_{U} - 1.5) - 1.51, \\
U &= 0.04 (U - B) - 0.36(X_{U} - 1.5) + 18.61 + u, \\ 
B - V &= 1.29(b - v) - 0.10(X_{B} - 1.5) - 0.33, \\
B &= 0.18 (B - V) - 0.22(X_{B} - 1.5) + 20.29 + b, \\
V - R &= 0.85(v - r) - 0.04 (X_{V} - 1.5) - 0.05, \\
V &= -0.14(V - R) - 0.14(X_{V} - 1.5) + 20.57 + v, \\
R-I &= 0.85( r - i ) - 0.05(X_{R} - 1.5) + 0.76, \\
R &= -0.04(R - I) - 0.10(X_{R} - 1.5) + 20.64 + r, \\
\end{align*}

where $U, B, V, R, I$ and $u, b, v, r, i$ are the calibrated and instrumental magnitudes, respectively. $X_{U}, X_{B}, X_{V}, X_{R}$ are the airmasses in the $U,B,V$ and $R$-bands respectively.

\subsection{Las Cumbres Observatory Global Telescope (LCOGT)}\label{subsec:lcogt}

We also monitored Gaia24ccy with the LCOGT telescopes \citep{2013PASP..125.1031B}. We observed from 2025, February 02, to May 25, gathering a total of 157 frames with 39, 40, 40, and 38 frames in $u,\ g,\ r,$ and $i$-bands, respectively. We performed the aperture photometry on the frames already calibrated by the LCOGT pipeline. We performed differential photometry with a few reference stars in the field, utilizing their calibrated magnitudes from SkyMapper \citep{2024PASA...41...61O}.

\subsection{Thai Robotic Telescopes (TRTs)}\label{subsec:TRTs}
National Astronomical Research Institute of Thailand (NARIT) operates four 0.7m telescopes, two in each hemisphere: TRT-CTIO (Cerro Tololo Inter-American Observatory, Chile), TRT-SBO (Springbrook Observatory, Australia), TRT-GAO (Gaomeigu Observatory, China), and TRT-SRO (Sierra Remote Observatories, USA). From 2025 January to April, 
we performed $U,B,V,R,I$ photometric observations of Gaia24ccy with TRT-CTIO and TRT-SBO. At each epoch, at least 3 frames were collected per filter, with an exposure time of 260s ($U$), 120s ($B$), 60s ($V$), 20s ($R$ and $I$). At TRT-CTIO, we collected 232, 231, 231, 227, and 227 frames in $U, B, V, R,$ and $I$ filters, respectively. At TRT-SRO, we collected 64 frames in $B$ and 67 in each of $V, R,$ and $I$-bands. The TRT pipeline produces science-ready World Coordinate System (WCS) calibrated frames after applying dark, bias, and flat corrections. A few filters on certain epochs were uncalibrated raw frames. We calibrated them with the concurrent calibration frames and Astrometry.net. Aperture photometry is calibrated with secondary standards in the field of view.

\subsection{All Sky survey light curves}\label{subsec:allskysurveydata}
We also used multiband light curves from several surveys. These include the $g$ and $V$-band light curves from All-Sky Automated Survey for Supernovae \citep[ASAS-SN,][]{2014ApJ...788...48S,2017PASP..129j4502K}, $B, V, R$ and $I$-band light curves from The American Association of Variable Star Observers (AAVSO), a high cadence optical light curve from K2 \citep{2018AJ....155..196R}, $zr$ and $zg$-band light curves from Zwicky Transient Factory \citep[ZTF,][]{2019PASP..131a8002B}. Additionally, we also incorporated $o$-band light curve from Asteroid Terrestrial-impact Last Alert System \citep[ATLAS,][]{2018PASP..130f4505T}, $W1$(3.4 $\mu$m) and $W2$(4.6 $\mu$m)-band light curves from the Near-Earth Object Wide-field Infrared Survey Explorer \citep[NEOWISE,][]{2011ApJ...731...53M}, and $G$-band light curve from Gaia \citep{2023A&A...674A...1G}.

\begin{deluxetable}{cccc}
    \tabletypesize{\footnotesize}
    \tablecaption{The HFOSC/HCT spectroscopic logs of Gaia24ccy. SNR is estimated at 5700 \r{A}.}\label{table:obs_log}
    \tablewidth{1pt}
    \tablehead{
    \colhead{JD} & \colhead{Exp (s)} & \colhead{Grating/Slit} & \colhead{SNR}
    }
    \startdata
    2460501.19 & 600 & Gr8/167l & 21 \\
    2460505.14 & 600,600 & Gr8, Gr14/167l, 167l &  20 \\
    2460515.21 & 720,720 & Gr8, Gr14/167l, 167l & 22 \\
    2460517.16 & 720,720 & Gr8, Gr14/167l, 167l & 20 \\
    2460740.47 & 1200 & Gr8/167l & 30 \\
    2460741.48 & 1800 & Gr8/167l & 38 \\
    2460742.47 & 1800 & Gr8/167l & 50 \\
    2460743.48 & 1800 & Gr8/167l & 46 \\
    2460872.15 & 1200 & Gr8/167l & 10 \\
    \enddata
\end{deluxetable}

\section{Analysis and Results}\label{sec:analysis}

In this section, we estimate the intrinsic parameters of Gaia24ccy, followed by an analysis of the outburst light curve profiles.

Gaia24ccy is a system of two close-by, similar NIR brightness YSOs, Gaia24ccy\,A and Gaia24ccy\,B. We performed Lomb-Scargle periodogram analysis on the K2 light curve (Appendix \ref{app:K2_period}) and identified two prominent peaks in its power spectra. We attributed the $P_{A}$=1.1419-day period to Gaia24ccy\,A and the $P_{B}$=1.7898-day period to Gaia24ccy\,B (Figure \ref{fig:KeckAO}). Furthermore, Gaia24ccy\,B's period prominently appeared just before the two outbursts in 2019 and 2024 (Appendix \ref{app:asasn_period}), suggesting it to be the outbursting object. In the subsequent analysis, we therefore assume that Gaia24ccy\,A remained in a quiescent state, while Gaia24ccy\,B underwent both outbursts.

\subsection{Stellar Parameters}\label{subsec:stellarparams}

Gaia24ccy\,A and Gaia24ccy\,B have nearly identical NIR brightness, $\Delta K=$ 0.26 mag \citep{2019ApJ...878...45B}. In the absence of other information about the individual objects, we assume that both objects are intrinsically similar during quiescence (see Section \ref{subsec:YSOsnature} and Appendix \ref{app:dipper}). 

The earlier estimates of the stellar mass treated Gaia24ccy as a single object. In these studies, the stellar mass was estimated either by comparing the observed spectral flux with the photospheric templates or by deriving the stellar parameters from the photometric fluxes \citep{2016ApJ...827..142B,2020A&A...639A..58M,2022A&A...663A..98T}. 
Because the system consists of two objects, these studies overestimated the masses of individual objects: Gaia24ccy\,A and Gaia24ccy\,B. 

However, \citet{2023ApJ...945..112F} performed $\chi^{2}$ minimization of the normalized source spectrum with normalized photospheric templates. This method depends solely on the stellar temperature and its gravity. Under our assumption of similarity of both objects, \citet{2023ApJ...945..112F} derived the spectral type (SpT) of a single object and thus, Gaia24ccy\,A and Gaia24ccy\,B are M3.8 type stars. 

We converted the SpT to temperature by interpolating the M1-M6 temperatures from \citet{2017AJ....153..188F}, yielding $T_\star$=3198 K for both objects. \citet{2023ApJ...945..112F} calculated the stellar luminosity from I$_{7500}$\footnote{Stellar intensity at 7500\r{A}, calculated from the flux calibrated spectrum. The choice of wavelength avoids a significant contribution from the excess emission of the accretion hotspot.}, using bolometric corrections from \citet{2014ApJ...786...97H}. The stellar luminosity was $L_{bol 2stars}$=10$^{-0.89}L_\odot$ = 0.129$L_\odot$\footnote{$L_{bol 2stars}$ is the bolometric luminosity with equal contributions from both the objects.}. The authors then estimated the stellar mass using the nonmagnetic pre-main-sequence tracks from \citet{2016A&A...593A..99F}, together with the stellar temperature and luminosity. The resulting stellar mass was $M_\star$=0.20$M_\odot$\footnote{We note that the luminosity adopted by authors is overestimated by a factor of two. Since the evolutionary tracks from \citet{2016A&A...593A..99F} tightly constrain the mass of late-type stars with their photospheric temperatures, we place confidence in the stellar mass estimated by \citet{2023ApJ...945..112F}.}. Finally, we estimated the stellar radius ($R_\star$) using the Stefan-Boltzmann's relation, \(L_{bol} = \sigma T_\star^{4}(4\pi R_\star^{2}) \), where $\sigma$ is the Stefan-Boltzmann constant and $L_{bol}$ is the bolometric luminosity of a single object. With \(L_{bol}= L_{bol2stars} /2 = 0.0644 L_\odot \), we obtained $R_\star$=0.83$R_\odot$.

We estimated the stellar parameters assuming both objects to be identical during quiescence, though this assumption may not hold. Future spatially resolved observations of Gaia24ccy\,A and Gaia24ccy\,B will be essential for a more robust characterization of the system.

\subsection{Mass Accretion rate}\label{subsec:massaccretionrate}

The mass-accretion rate is reliably estimated by fitting the Balmer jump of the spectrum \citep{1998ApJ...509..802C,2014PhDT.......477M}. Alternatively, empirical relations have been established to estimate the accretion luminosities from specific spectral line fluxes and $U$-band excess over the stellar photosphere \citep{1998ApJ...492..323G,2014A&A...570A..82V,2017A&A...600A..20A,2018ApJ...868...28F}.

\citet{2020A&A...639A..58M} estimated quiescent state mass-accretion rate ($\dot{M}_{qui}$) of Gaia24ccy to be 2.93 $\times$ 10$^{-9}$ $M_\odot\,\mathrm{yr}^{-1}$ by fitting the dereddened X-shooter spectrum with a sum of a hydrogen slab and class II photospheric templates. \citet{2023ApJ...945..112F} estimated $\dot{M}_{qui}$ = 4.68 $\times$ 10$^{-9} M_\odot\,\mathrm{yr}^{-1}$ by calculating the accretion luminosities ($L_{\rm acc}$) from the spectral line luminosities (H$\alpha$, H$\beta$, HeI$\ \lambda\lambda\ 5876\  \&\ 6678$ \r{A}) following relations from \citet{2018ApJ...868...28F}\footnote{The relation coefficients in \citet{2018ApJ...868...28F} are slightly different from those in \citet{2017A&A...600A..20A}, but mostly within the 1$\sigma$ error.}. Under the aforementioned assumption of similarity of Gaia24ccy\,A (see Section \ref{subsec:YSOsnature} and Appendix \ref{app:dipper}) and Gaia24ccy\,B, the quiescent mass-accretion rate of each of the sources is 2.34 $\times$ 10$^{-9} M_\odot\,\mathrm{yr}^{-1}$ \citep{2023ApJ...945..112F}.

\begin{figure*}
    \centering
    \includegraphics[width=1\linewidth]{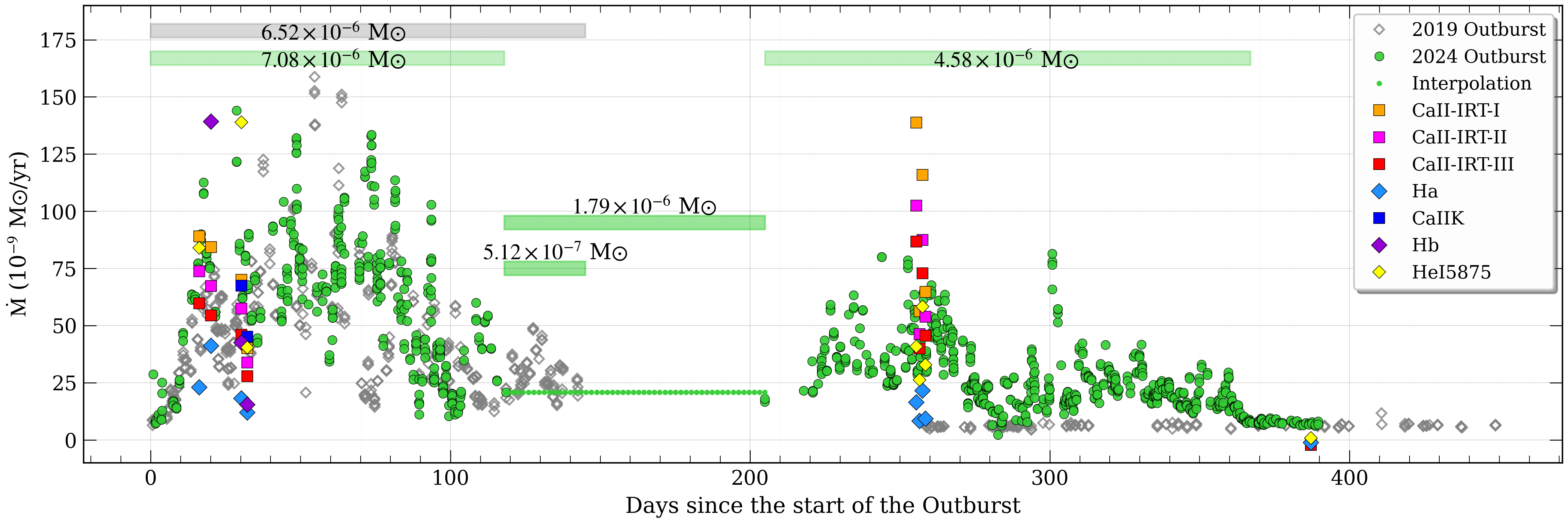}
    \caption{The accretion rate profiles during the two outbursts of Gaia24ccy\,B. Grey diamonds cover the 2019 outburst, while green circles cover the 2024 outburst. The 2019 accretion profile is estimated from the ASAS-SN $g$ light curve, while the 2024 profile is estimated from ASAS-SN $g$, TMMT $U$, LCOGT $u$, and AAVSO $B$ light curves. The two halves of the 2024 outburst, before 118 days and after 205 days, are connected with linear interpolation shown in green dots. Grey and green shaded strips highlight a time interval, and the respective texts are the total mass accreted in that time interval. The colored squares and pentagons represent mass-accretion rates estimated from HFOSC spectral lines. The wavelengths of CaII-IRT-I/II/III are 8498, 8542, and 8662 \r{A}, respectively.}
    \label{fig:MassAcc_replicaOutburst}
\end{figure*}

Similarly, using our flux-calibrated HFOSC spectra, we calculated the luminosities of few spectral lines: He I 5875, H$\alpha$,\ H$\beta$,\ Ca\,II\,$\lambda\lambda\ 3933,\ 8498,\ 8542\ \&\ 8662$ \r{A} and estimated the $L_{\rm acc}$ following relations from \citet{2017A&A...600A..20A}. The spectra were dereddened for A$_{V}$ = 1.3 mag \citep{2023ApJ...945..112F}, using the extinction curve from \citet{1989ApJ...345..245C}. We estimated the mass-accretion rate ($\dot{M}_{\rm acc}$) by equating the accretion luminosity to the release of gravitational energy per unit time as matter falls from a certain radius ($R_{in}$)\footnote{At inner disk radius ($R_{in}$), the disk is truncated by the stellar magnetic field by an equivalence of disk ram pressure and stellar magnetic pressure \citep{2007prpl.conf..479B}.} onto the stellar surface,

\begin{equation}\label{equ:massaccrate}
    \dot{M}_{\rm acc} = \frac{L_{\rm acc}R_\star}{GM_\star} \left( 1 - \frac{R_\star}{R_{in}}  \right)^{-1},
\end{equation}

\noindent where $G$ is the gravitational constant. $M_\star$ and $R_\star$ are already defined. The inner disk radius ($R_{in}$) is typically 5$R_\star$ \citep{1998ApJ...492..323G,2008ApJ...681..594H}. These estimates include contributions from both the objects, Gaia24ccy\,A and Gaia24ccy\,B. To trace the evolution of Gaia24ccy\,B’s accretion rate along the outburst, we subtracted the quiescent accretion rate of Gaia24ccy\,A from the total mass accretion rate. The accretion rates of Gaia24ccy\,B are shown in colored squares and rhombuses in Figure \ref{fig:MassAcc_replicaOutburst} and are also tabulated in Table \ref{table:massaccrate}. The mass-accretion rates derived from seven spectral lines show significant variation at each epoch (see Figure \ref{fig:MassAcc_replicaOutburst}). The mass-accretion rates from H$\alpha$ are consistently lower than those from photometry and other spectral lines. The mass-accretion rate measurements from He I 5875 \r{A} are consistent with the photometric estimates, except for the second epoch. Ca\,II\,IR triplet (Ca\,II\,IRT) lines also show a good match to the photometric measurements of $\dot{M}_{\rm acc}$, except on 2025 March 05 and 07. The quiescent accretion rate of Gaia24ccy\,B, estimated from the He I line in 2025 July 15 spectrum, is $0.87 \times 10^{-9} M_\odot\,\mathrm{yr}^{-1}$, which is 2.7 times lower than that from \citet{2023ApJ...945..112F}. However, this is consistent with the spread in the photometric mass-accretion rates during the quiescence, which spans a factor of $\sim$2.2. All other spectral lines yield a total accretion rate of $1.2 \times 10^{-9} M_\odot\,\mathrm{yr}^{-1}$ for the Gaia24ccy system, implying a negative quiescent accretion rate for Gaia24ccy\,B. This could indicate Hydrogen and Ca II lines originating from different regions.

\begin{deluxetable*}{ccccccccc}
\tablecaption{The mass-accretion rate of Gaia24ccy\,B from HFOSC spectral lines. It is in units of 10$^{-9} M_\odot\,\mathrm{yr}^{-1}$. We do not show negative mass-accretion rate from the last epoch, except for He I 5875 \r{A} line.  }\label{table:massaccrate}
\tablewidth{1pt}
\tablehead{
\colhead{JD} & \colhead{He I} & \colhead{H$\alpha$} &
\colhead{H$\beta$} & \colhead{Ca\ II\ K} & \colhead{Ca\ II\ IRT\ I} & \colhead{Ca\ II\ IRT\ II} & \colhead{Ca\ II\ IRT\ III} \\
\colhead{} & \colhead{5875 \r{A}} & \colhead{6563 \r{A}} & \colhead{4861 \r{A}} & \colhead{3934 \r{A}} & \colhead{8498 \r{A}} & \colhead{8542 \r{A}} & \colhead{8662 \r{A}}
}
\startdata
2460501.19 & 84.1 $\pm$ 49.9 & 23.0 $\pm$ 12.9 & ---             & ---              & 89.0 $\pm$ 68.1  & 73.7 $\pm$ 58.9 & 59.7 $\pm$ 49.7 \\
2460505.14 & 224.7 $\pm$ 127.2 & 41.1 $\pm$ 21.6 & 139.2$\pm$ 60.4 & 566.9$\pm$ 256.2 & 84.3 $\pm$ 64.7  & 67.2 $\pm$ 54.2 & 54.4 $\pm$ 45.6 \\
2460515.21 & 138.9 $\pm$ 80.3 & 18.1 $\pm$ 10.5 & 42.6 $\pm$ 20.1 & 67.4 $\pm$ 34.0  & 70.0 $\pm$ 54.3  & 57.4 $\pm$ 46.8 & 46.0 $\pm$ 39.1 \\
2460517.16 & 40.4 $\pm$ 25.3 & 11.9 $\pm$ 7.5  & 15.3 $\pm$ 8.3  & 45.2 $\pm$ 23.5  & 40.2 $\pm$ 32.5  & 33.8 $\pm$ 28.9 & 27.8 $\pm$ 24.9 \\
2460740.47 & 40.8 $\pm$ 25.5 & 16.4 $\pm$ 9.6  & ---             & ---              & 138.8$\pm$ 104.0 & 102.5$\pm$ 80.2 & 86.7 $\pm$ 70.3 \\
2460741.48 & 26.3 $\pm$ 17.2 & 8.3  $\pm$ 5.6  & ---             & ---              & 56.3 $\pm$ 44.4  & 46.2 $\pm$ 38.4 & 40.0 $\pm$ 34.5 \\
2460742.47 & 58.3 $\pm$ 35.5 & 21.6 $\pm$ 12.2 & ---             & ---              & 115.9$\pm$ 87.5  & 87.5 $\pm$ 69.2 & 72.9 $\pm$ 59.8 \\
2460743.48 & 32.9 $\pm$ 21.0 & 9.2  $\pm$ 6.1  & ---             & ---              & 64.7 $\pm$ 50.5  & 53.9 $\pm$ 44.1 & 45.6 $\pm$ 38.8 \\
2460872.15 & 0.87 $\pm$ 2.1  & ---             & ---             & ---              & ---              & ---             & ---             \\
\enddata
\end{deluxetable*}

Along with enhanced spectral lines, accretion manifests dramatically in the blue portion of the spectrum by releasing a significant portion of the accretion energy as an excess over the photosphere, which is captured by the $U$-band \citep{1998ApJ...492..323G, 1998ApJ...509..802C}. This $U$-band excess luminosity ($L_{U-excess}$) exhibits a tight correlation with the accretion luminosity \citep{1998ApJ...492..323G,2014A&A...570A..82V,2018ApJ...852...56G,2023ApJ...957..113W}. We utilized our TMMT $U$-band observations during the peak of the outburst to calculate the accretion luminosity. The TMMT $U$-band magnitudes were converted to the flux unit, using zero magnitude absolute flux provided by \citet{1998A&A...333..231B}, and were extinction corrected for $A_{V}=1.3$ mag \citep{2023ApJ...945..112F}. We estimated the $U$-band excess flux by subtracting the photospheric and chromospheric contribution using an M4 spectral template of a Weak-line T Tauri Star (WTTS), SZ 94, with a photospheric temperature of 3190 K \citep{2011ApJ...743..105I,2013A&A...551A.107M,2017A&A...605A..86M,2024A&A...690A.122C}. The template was reprojected to the distance of Gaia24ccy, 142 pc. Similarly, we estimated excess flux in ASAS-SN $g$-band and developed a correlation with $U$-band excess fluxes to access high-cadence and long-term measurements of the mass-accretion rate\footnote{Assuming that the YSO is a sum of two blackbodies: photosphere and hotspot. Eliminating the photospheric contribution, we are left with one blackbody. The two wavelengths of the same blackbody would produce a better-constrained linear relation. If an increase in mass-accretion rate scales with the hotspot area, we would see the same linear relation between $U_{excess}- g,B_{excess}$ at various states of accretion. The relation should change if the accretion rate variation causes thermal variations in the hotspot.}. We also established a linear relation between AAVSO $B$ and TMMT $U$-band excess fluxes (see Figure \ref{fig:Uexcess_gBexcess}). The dereddened photometry selected for establishing these correlations was observed within 2.5 hours. The coefficients did not change significantly for data points as close as 1 hour or 0.5 hour. The relations are:

\begin{equation}\label{equ:Uexcess_gBexcess}
\begin{aligned}
    F_{U-excess} = 1.095(\pm0.102)F_{B-excess} + 0.17(\pm0.066), \\
    F_{U-excess} = 1.088(\pm0.063)F_{g-excess} + 0.113(\pm0.065), 
\end{aligned}
\end{equation}

\noindent where, $F_{i-excess}$ is the excess flux in the $i$-band over the stellar photosphere. $F_{i-excess}$ and intercepts are in units of $10^{-13}\ \mathrm{erg\,s^{-1}\,cm^{-2}}$\r{A}$^{-1}$. The excess luminosity in $U$-band was then estimated following \citet{2023ApJ...957..113W}, 

\begin{equation}\label{equ:UexcessLum}
    L_{U-excess} = 4\pi d_{\star}^{2} \times \Delta\lambda_{U} \times (F_{U-excess})
\end{equation}

\begin{figure}\label{fig:Uexcess_gBexcess}
    \centering
    \includegraphics[width=1\linewidth]{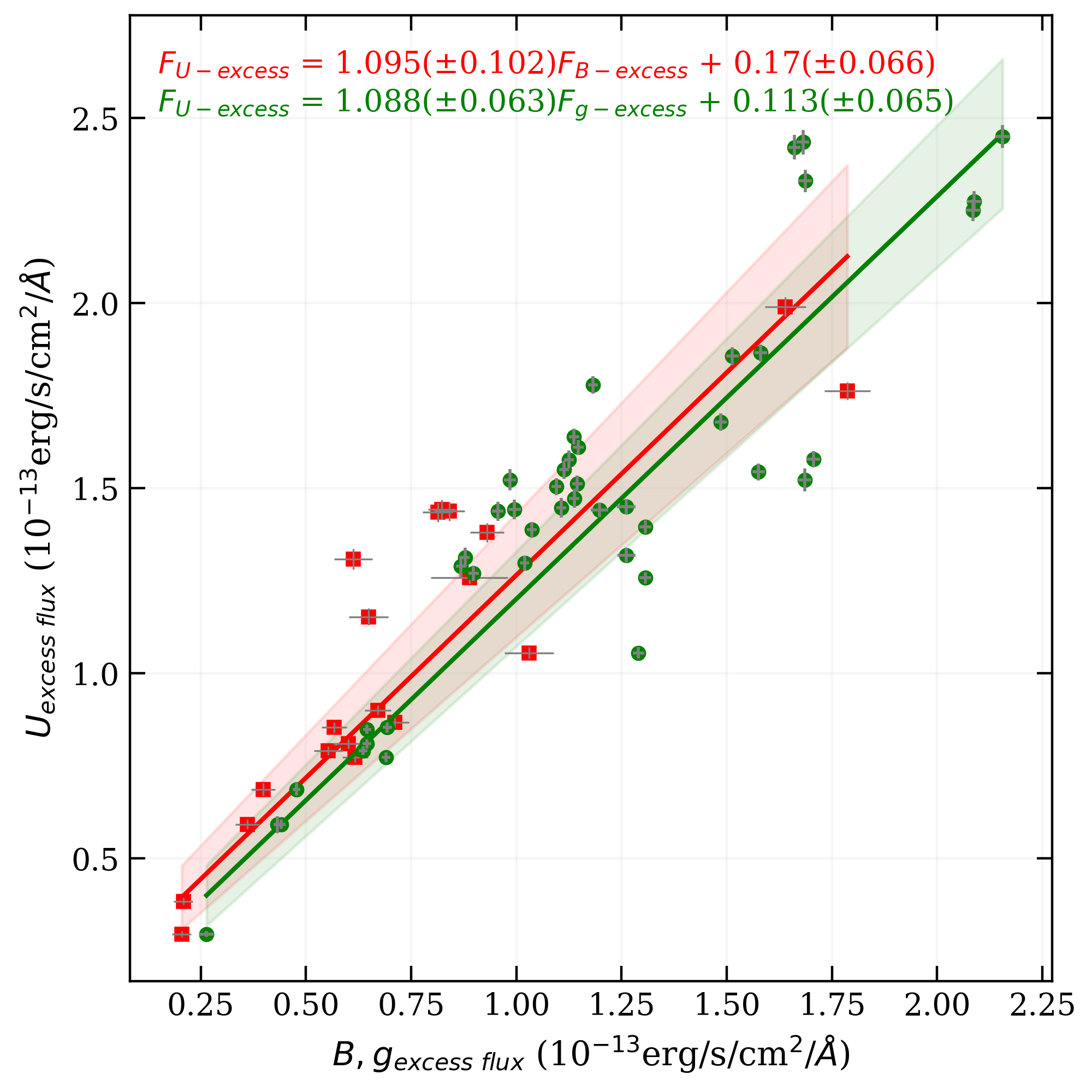}
    \caption{The linear relations between the $F_{U-excess}$ and $F_{B-excess}, F_{g-excess}$ during the peak of the outburst are shown in dark red and green lines with 1-sigma error in light shades (Equation \ref{equ:Uexcess_gBexcess}). $F_{i-excess}$ and intercepts are in units of $10^{-13}\ \mathrm{erg\,s^{-1}\,cm^{-2}}$\r{A}$^{-1}$.}
    \label{fig:Uexcess_gBexcess}
\end{figure}

where $d_{\star}$=142 pc is the distance to Gaia24ccy and $\Delta\lambda_{U}$ $\sim$680 \r{A} is the bandwidth of $U$ filter. We estimated $L_{\rm acc}$ from $L_{U-excess}$ using the correlation in \citet{1998ApJ...492..323G} and then translated it to mass-accretion rate with Equation \ref{equ:massaccrate}.

\citet{2014A&A...570A..82V} noted that the accretion rates estimated from $u$-band excess luminosities following the relation in \citet{2014A&A...570A..82V} were overestimated by a factor of $\leq 2-4$ from those of \citet{1998ApJ...492..323G}. We also tested this by calculating the mass-accretion rate from LCOGT $u$-band excess luminosities using \citet{2014A&A...570A..82V}; and estimated the scaling factor against those from Equations \ref{equ:Uexcess_gBexcess}, \ref{equ:UexcessLum} and \citet{1998ApJ...492..323G}. The relation is,

\begin{equation}\label{equ:u_acc_corr}
    \dot{M}_{acc\,-8} = 0.59(\pm 0.09) \dot{M}u_{-8} - 1.27(\pm 0.78)
\end{equation}

\noindent where $\dot{M}u_{-8}$ and $\dot{M}_{acc\,-8}$ are accretion rate estimates from LCOGT $u$-band, and equation \ref{equ:UexcessLum}, respectively. $\dot{M}u_{-8}$, $\dot{M}_{acc\,-8}$, and intercept are in units of 10$^{-8} M_\odot\,\mathrm{yr}^{-1}$. The scaling factor of 0.59 is consistent with \citet{2014A&A...570A..82V}.

The errors in mass-accretion rates increased up to $\sim100\%$. These errors are not independent for each observation epoch but are largely dominated by the systematic errors propagating in through the intercept of Equation \ref{equ:Uexcess_gBexcess}.

Figure \ref{fig:MassAcc_replicaOutburst} shows the mass accretion rate profile of Gaia24ccy\,B. The photometrically derived median mass accretion rate during the pre-outburst quiescence is 5.4 $\times$ 10$^{-9} M_\odot\,\mathrm{yr}^{-1}$, which is nearly twice the accretion rate estimated by \citet{2023ApJ...945..112F}, 2.3 $\times$ 10$^{-9} M_\odot\,\mathrm{yr}^{-1}$. This quiescent mass accretion rate appears overestimated, likely because the ASAS-SN $g-$TMMT $U$ correlation (Equation \ref{equ:Uexcess_gBexcess}) established during the outburst may not hold during the quiescence. The observation in \citet{2023ApJ...945..112F} was performed on 2013 June 4.

The post-outburst mass-accretion rate of Gaia24ccy system estimated from TMMT $U$-band excess luminosity (last cyan squares in the \textit{upper panel} of Figure \ref{fig:lightcurve}) is 2 $\times$ 10$^{-9} M_\odot\,\mathrm{yr}^{-1}$, with a spread of $0.4 - 5 \times$ 10$^{-9} M_\odot\,\mathrm{yr}^{-1}$, which is consistent with the estimates from \citet{2020A&A...639A..58M} and \citet{2023ApJ...945..112F}. This spread, however, contrasts with the pre-outburst quiescence, where it was only $\sim 2.2$.

The peak accretion rates of Gaia24ccy\,B during the 2019 and 2024 outbursts were 1.57 $\times$ 10$^{-7} M_\odot\,\mathrm{yr}^{-1}$ and 1.42 $\times$ 10$^{-7}M_\odot\,\mathrm{yr}^{-1}$, respectively. These peak accretion rates are $\sim$67 and $\sim$61 times the quiescence state value. This suggests that a naive conversion of ratio of photometric flux rise (33$-$36 $\equiv \Delta g$ = 3.8$-$3.9 mag) would have underestimated the increase in accretion rate of the outbursting component, Gaia24ccy\,B, had two object not been resolved.

\subsection{Light curves and outbursts}\label{subsec:lightcurves}

Figure \ref{fig:lightcurve} shows the multiband light curves of Gaia24ccy. Magnitude offset has been applied to the light curves for clarity. As discussed earlier, the optical light curves show two very similar $\Delta g =$ 3.8 mag outbursts in 2019 and 2024 (shown as orange shaded regions), separated by a long period of stable brightness. NEOWISE \textit{W1} and \textit{W2} light curves caught the 2019 outburst with a sharp rise. Interestingly, both \textit{W1} and \textit{W2} light curves show a backward rising trend around 2014, which coincides with a small, $\Delta V \sim$1 mag, photometric bump in the ASAS-SN light curve. This feature could be a missed outburst (with only the tail captured) or could be a regular variability, similar to optical brightening seen in ASAS-SN $g$-band light curve around 2022 June (Figure \ref{fig:lightcurve}). With this ambiguity, we call the 2014 brightening event a `2014 outburst?' (with a question mark). Outbursts recur at least twice with a $\sim$5 year interval; a third event in 2014 is a possibility, but unconfirmed. An outburst in 2029 will be critical for the confirmation of the 5 year periodicity.

\subsection{2024 outburst: A replica of 2019 outburst}\label{subsec:replicaOutburst}
The 2019 and 2024 outbursts were observed by ASAS-SN \textit{g}-band, providing a direct opportunity to compare the profiles that evolved similarly (see \textit{lower panel} in Figure \ref{fig:lightcurve}). Figure \ref{fig:MassAcc_replicaOutburst} shows the mass-accretion rate profiles, with the 2019 outburst in grey rhombuses and the 2024 outburst in green circles. 

The light curves during both the outbursts rose sharply for the initial 15 days, at a rate of 0.160 mag/day (2019) and 0.166 mag/day (2024). This rise was followed by a $\Delta g=$ 1.3 mag dip during 2019, while the coevolving 2024 dip is observationally incomplete. After rising from the dip, both light curves reached a mean magnitude of $\sim$13 and showed variability for the following 45 days. The 2019 outburst light curve underwent a sharp dip of $\Delta g=$2 mag at 75 days; however, a similar dip did not appear during the 2024 outburst. At 80 days, the light curves began to decline nearly linearly, and faded to $g=$ 14.4 mag by day 100. Then the 2019 outburst exhibited two clear $\Delta g \sim$ 1.5 mag `sub-bursts' (see Section \ref{subsec:subbursts}) before the end of the object's annual visibility.

In 2019, when Gaia24ccy\,B reappeared for its annual visibility window at 255 days, it had already returned to its pre-outburst quiescence level. However, in 2024, it did not return to quiescence for 367 days after the onset of the outburst. By the time the 2019 outburst had settled, the 2024 outburst was undergoing multiple $\Delta g \sim$1.2 mag sub-bursts.

\subsection{Sub-bursts during the 2019 and 2024 outburst}\label{subsec:subbursts}

\begin{figure}
    \centering
    \includegraphics[width=1\linewidth]{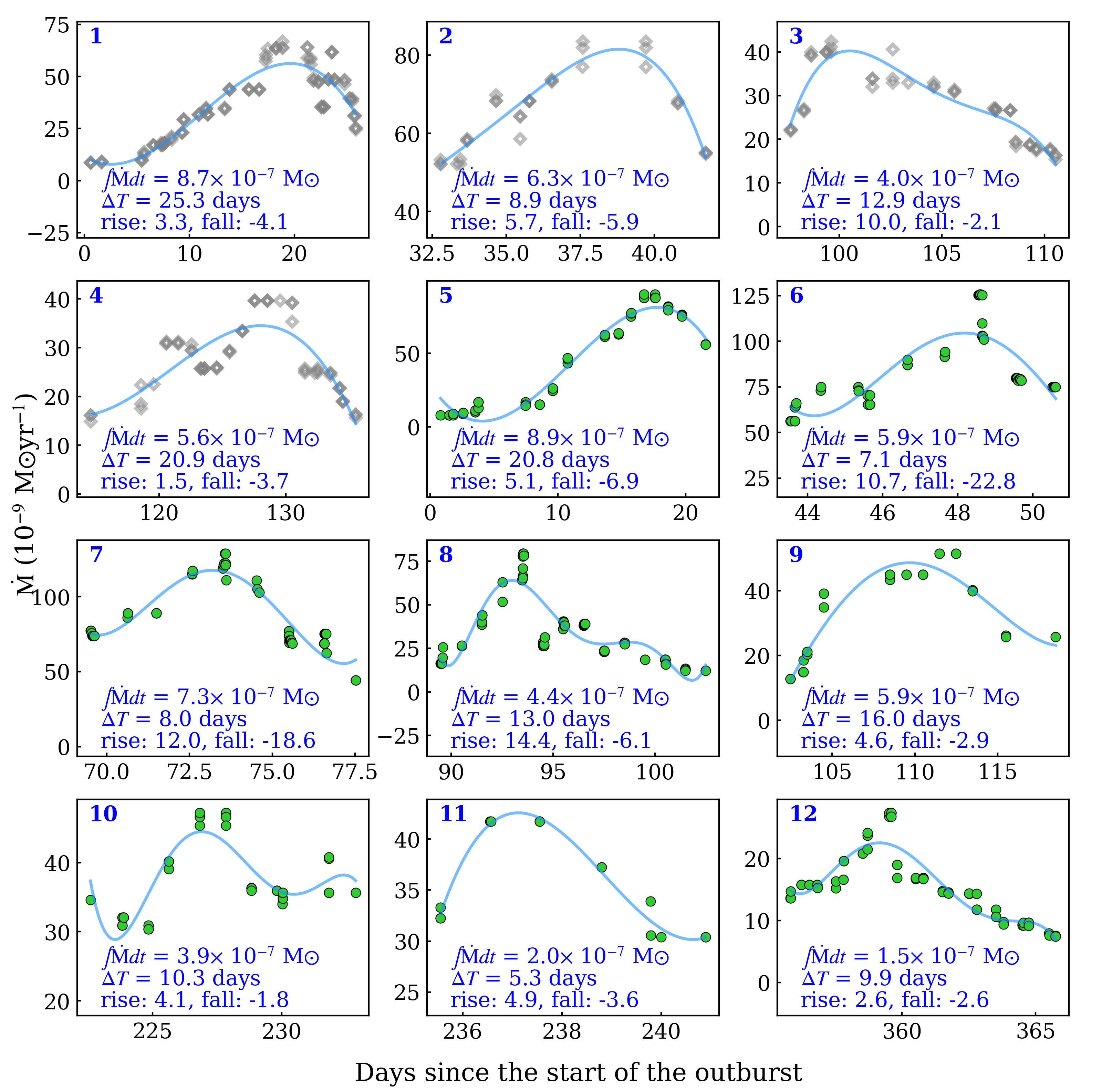}
    \caption{Sub-bursts during 2019 (grey rhombus in the 1$^{st}$ and 2$^{nd}$ rows) and 2024 (green circles in 2$^{nd}$, 3$^{rd}$ and 4$^{th}$ rows) are shown. A polynomial fit of 3-6 degrees is shown as a light blue curve. The timescale of the sub-burst ($\Delta T$),  the total mass accreted ($\int {\dot{\text{M}}dt}$), and slopes of rise and fall (in units of 10$^{-9} M_\odot\,\mathrm{yr}^{-1}\mathrm{day}^{-1}$) are written in the respective panels. Each panel is marked with a blue integer written at the upper left corner.}
    \label{fig:subbursts}
\end{figure}

During the 2019 and 2024 outbursts, several $\Delta g \sim$0.5 - 1.5 mag photometric bumps appeared over a timescale of 5 to 25 days, with mass accretion rate increasing up to a factor of 10 (see Figures \ref{fig:MassAcc_replicaOutburst} and \ref{fig:subbursts}). These 5 to 25-day photometric bumps differ from the few-hour timescale clumps reported by \citet{1996A&A...307..791G, 2024ApJ...968...88S}. They are also longer than the rotation period at the corotation radius, that is, the radius where the Keplerian period of the disk is the same as the stellar period. This suggests that the events are related to neither the free fall along the accretion channel nor the magnetosphere at the truncation radius, but are regulated by the disk physics. The mechanism triggering these events could be similar to the mechanism regulating accretion outbursts. We call these events `Sub-bursts'. We also termed the initial outburst rise, followed by a dip, as a sub-burst (panels 1 and 5 in Figure \ref{fig:subbursts} for 2019 and 2024 outbursts, respectively).

The total mass accreted during each sub-burst is written in the corresponding panels of the Figure \ref{fig:subbursts}. These masses could be underestimated because parts of the events were often missed observationally. The sub-bursts exhibit different morphologies: slow rise and fast decay, fast rise and slow decay, and similar rates of rise and decay. The rates of rise and fall (in units of 10$^{-9} M_\odot\,\mathrm{yr}^{-1}\mathrm{day}^{-1}$) are written in the respective panels. \citet{2024MNRAS.530.1749N} found similar `sub-bursts' in the decaying mass-accretion rate profile of thermal-instability driven outburst. The authors called them `reflares'.

\subsection{Color-evolution: Optical Blueing and MIR reddening during the Outburst}\label{subsec:blueing_reddening}

\begin{figure}
    \centering
    \includegraphics[width=1\linewidth]{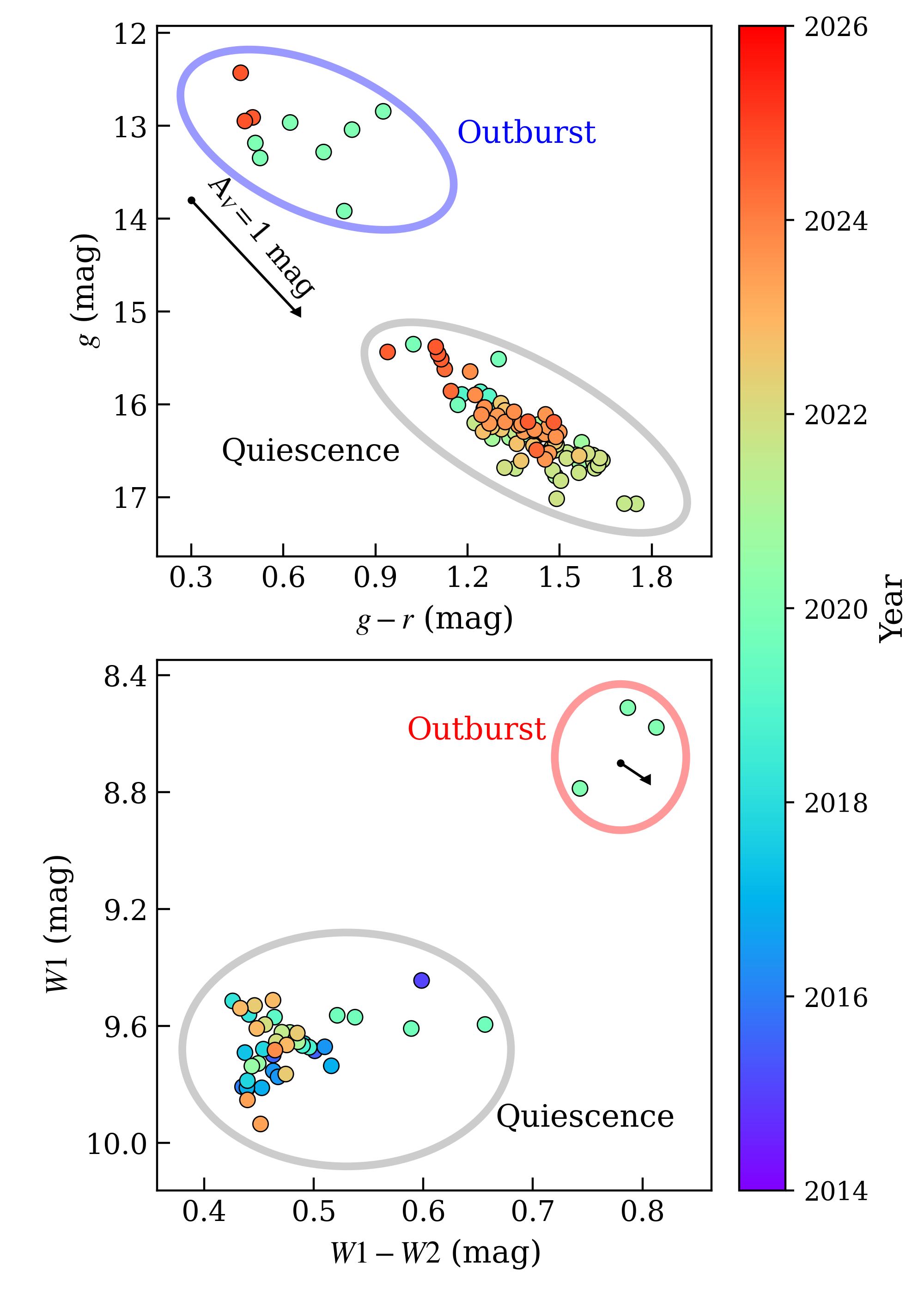}
    \caption{\textit{upper}: Optical color-magnitude plot is shown with ZTF $g$ and $r$-band light curves. The colors of the dots represent the epoch of observation. Ellipses separately encircle the quiescence and outburst colored dots. The lime and red color dots within the outburst ellipse represent the 2019 and 2024 outbursts, respectively. The optical color became bluer during the outburst. \textit{lower}: MIR color-magnitude plot is shown with NEOWISE $W1$ and $W2$-band light curves. The MIR color becomes redder during the outburst. An extinction vector for A$_{V}$=1 mag is shown.}
    \label{fig:color-mag}
\end{figure}

The NEOWISE coverage of the 2019 outburst, together with the similar optical outburst profiles of 2019 and 2024, provides an opportunity to analyze the color evolution across both outbursts. Figure \ref{fig:color-mag} shows the optical ZTF ($g-r$) and MIR NEOWISE ($W1-W2$) color magnitude diagram across the outbursts. The optical color followed expectations of an enhanced accretion-driven event, becoming bluer as the source became photometrically brighter during both outbursts. In contrast, the MIR color reddened significantly during the 2019 outburst. Along with this secular `blue-ing' and reddening, Gaia24ccy\,B experienced regular color variation, as highlighted by the encircled points. Similar `reddening during brightening' has also been reported earlier by \citet{2021ApJ...920..132P,2025MNRAS.tmp..107M}. Both the optical and MIR colors exhibit a significant fluctuation. During quiescence, the median optical color was $g-r$ = 1.42 mag, with a spread between 0.94 - 1.75 mag. Similarly, the median optical color during the outburst was $g-r$ = 0.57 mag (range: 0.93-0.46 mag). Similar estimates in MIR color during the quiescence and outburst were 0.46 mag (range: 0.66-0.43 mag) and 0.79 mag (range: 0.81-0.74 mag). Thus, the color change during the outburst was $\Delta(g-r)$ = $-$0.85 mag and $\Delta(W1-W2)$ = 0.33 mag.

\section{Discussion}\label{sec:discussion}

\subsection{Nature of the two objects: Both are accreting YSOs}\label{subsec:YSOsnature}
Understanding the physical nature of Gaia24ccy\,A and Gaia24ccy\,B is crucial for interpreting key observables such as their periods, mass-accretion rates, and outburst triggers.

The NIRC2 AO image (Figure \ref{fig:KeckAO}) revealed two nearby objects at a projected distance of 19.57 au. Both objects appeared as fast rotators in the power spectrum of the K2 light curve (Appendix \ref{app:K2_period}). The Lomb-Scargle periodogram analysis of the ASAS-SN light curves showed that the period of Gaia24ccy\,B became prominent in the observation windows \textbf{f} (just before the 2019 outburst, see Appendix \ref{app:asasn_period} and Figure \ref{fig:ASASN_periodogram}), and \textbf{k} (just before the 2024 outburst). The appearance of Gaia24ccy\,B's period just before the outburst links the object directly to the event. Moreover, the median and peak-to-peak ASAS-SN $g$ flux level rose in these two windows, with respect to the preceding quiescent state windows, \textbf{e} and \textbf{j}. The simultaneous rise of peak-to-peak and mean fluxes suggests that the hotspot contribution from Gaia24ccy\,B to the total flux could have been amplified. The photometric rise (and so enhanced accretion rate), even before the actual onset of the outburst, indicates that the outbursting object is Gaia24ccy\,B. Since Gaia24ccy\,B underwent two outbursts, it is an accreting object and hence a YSO. 

The dips in the K2 light curve (see Appendix \ref{app:dipper}) provide insights into the nature of the second object. Appendix \ref{app:dipper} suggests that the dips could possibly result from quasi-periodic occultations of Gaia24ccy\,A.  Such dips have been reported earlier and are well studied \citep[see, for example,][]{1999A&A...349..619B}. Previous studies suggest two main explanations of the dips: (i) the occultation of the central star by warps in the disk material near the corotation radius \citep{1999A&A...349..619B}, and (ii) by the dust material from the disk rising along the magnetosphere, which periodically occults the star as we face the accretion funnel \citep{2024A&A...688A..61N} (or hotspot if there is no azimuthal twisting of the funnel). Both cases require the object to have a surrounding highly inclined disk, enabling the occultation of the central object. This suggests the possibility of Gaia24ccy\,A harboring a circumstellar disk and undergoing accretion. Thus, we assume, for this study, that Gaia24ccy\,A is also a YSO with a highly inclined disk.

\subsection{On Outburst Triggering Mechanism}\label{subsec:outburstMech}

Figure \ref{fig:MassAcc_replicaOutburst} shows that the outbursts' profiles remained similar for at least 100 days after onset. Comparable timescales and flux scales of the events suggest that the initial disk conditions leading to the outbursts could have been remarkably similar, even after 5 years. The repeatability of the outburst rules out the irregular external triggers, such as infall of accretion streams onto the disk or the stellar flybys \citep{2018MNRAS.475.2642K,2022MNRAS.510L..37B}.

In a possible binary system, Gaia24ccy\,A could trigger a gravitational instability in the inner disk of Gaia24ccy\,B at periastron, initiating an accretion outburst \citep{2017ApJ...842L..12T,2023EPJP..138...11C}. However, if the two objects are in a circular orbit with a radius of the current projected separation of 19.57 au, Kepler's 3$^{rd}$ law of motion suggests an orbital period of $\approx$137 years. This suggests the objects would move by only 13$^{\degree}$ over the 5 years, and so, this trigger mechanism seems unlikely. 

Among regular external perturbations, a planet in an eccentric orbit of period $\sim5$ years could gravitationally perturb the disk at periastron. Such a planet would be in an orbit with a radius of 1.71 au.

Two intrinsic mechanisms are, however, more natural explanations for repeated bursts in a replenishable inner disk and are discussed below. The first is a thermal–viscous instability (TI), in which a local heating–cooling imbalance (often aided by mass pile-up or a change in the effective viscosity) drives a disk annulus to a hot, high–viscosity state and produces an outburst \citep{1994ApJ...427..987B}. The second is magnetospheric gated-accretion (the D’Angelo–Spruit \citep{2012MNRAS.420..416D} or DS mechanism), where material accumulates at the magnetospheric truncation radius ($R_{T}$) and is released episodically onto the star. Below, we treat TI and magnetospheric/DS scenarios as distinct, testable possibilities.

We use ‘initial conditions’ in two distinct senses. For the TI scenario, this means the local thermal/viscous state (temperature, surface density, ionization) at the unstable radius. For the DS/gated-accretion scenario it refers to the accumulation (pile-up) of mass and pressure at the magnetospheric boundary. Below we treat these two meanings separately and present the diagnostics appropriate to each.

\subsubsection{Reset Mechanism}\label{subsubsec:Reset_mechanism}

The remarkable resemblance between the two outburst profiles and the same rate of rise for the initial 15 days suggests that the outbursts could have been triggered by similar initial conditions at the same disk radius, on both occasions. The two similar outbursts are separated by $\approx$5 yr; this recurrence could reflect either a characteristic replenishment timescale in the outer disk or be coincidental. We cannot establish periodicity from two events. Therefore, below we consider multiple physical scenarios that can produce similar repeated bursts. 

Figure \ref{fig:MassAcc_replicaOutburst} shows the amount of accreted matter during the outburst. In the 2019 outburst, Gaia24ccy\,B accreted $6.52 \times 10^{-6} M_\odot$ within 145 days after the onset of the outburst. By 255 days after the outburst, Gaia24ccy\,B had returned to quiescence. Similarly, $7.08 \times 10^{-6} M_\odot$ amount of matter was accreted in the initial 120 days during the 2024 outburst. To compare the accretion in the similar duration of 2019 and 2024 outbursts, we linearly interpolated the accretion rate profile of the 2024 outburst (see green dots in Figure \ref{fig:MassAcc_replicaOutburst}, from day 118 to 205). Accounting for the additional 25 days from the linear interpolation, the total accreted material during the 2024 outburst is estimated as 7.59 $\times 10^{-6} M_\odot$. Instead of our linear interpolation, an assumed sub-burst in those additional 25 days could have added 8 - 9 $\times 10^{-7}M_\odot$ (maximum mass in Figure \ref{fig:subbursts}). This will amount to a total of 7.88 - 7.98 $\times 10^{-6}M_\odot$ accreted mass during the first 145 days of 2024 outburst. This suggests that a similar amount of matter was accreted onto the star during the initial 145 days of both outbursts, with the slight difference attributed to the observation cadence. This accreted mass represents the lower limit of critical mass ($M_{\rm critical}$) (corresponding to the observed limiting duration of the 2019 outburst) required for the outburst to trigger, such that, \\

$M_{\rm critical} = $ 7.6 $\times 10^{-6} M_{\odot}$ \\

Accumulation of 7.6 $\times$ 10$^{-6} M_\odot$ mass would have reset the initial conditions that triggered the outburst at a certain radius in the disk, $r_{\rm trigger}$.

\subsection{Timescales, mass budget and diagnostics for TI vs DS}\label{subsec:TI_DS}

Orbital periods and Keplerian frequencies in the very inner disk are short: for example, the Keplerian period is $\sim0.3$ days at $r\simeq1.3R_\star$, $\sim2.2$ days at $r\simeq5R_\star$, $\sim2.9$ days at $r\simeq6R_\star$, and $\sim14.4$ days at $r\simeq17.2R_\star$. These short dynamical times ($1/\Omega(r); \text{where}\ \Omega(r) = \sqrt{GM\star /r^{3}}$ is the angular speed) set the minimum possible rise times for any process that operates at small radii.

\subsubsection{Timescales: definitions and diagnostic use}\label{subsubsec:TIDS_timescales}

For the purposes of diagnosing the outburst mechanism, we distinguish two characteristic timescales. The decay of an outburst is governed by the hot-state viscous time at the outer edge of the unstable region \citep{2023ASPC..534..355F},
\begin{equation}\label{equ:viscoustimescale}
    t_\nu(r)=\frac{r^{3/2}}{\alpha h^2\sqrt{GM_\star}},\qquad h\equiv\frac{H}{r},
\end{equation}
where $\alpha$ is a dimensionless viscosity parameter and \textit{H} is the disk scale height at disk radius $r$. The rise is set by the (local) thermal time,
\begin{equation}\label{equ:thermaltimescale}
    t_{\rm th}(r)=\frac{1}{\alpha\Omega(r)}.
\end{equation}
Eliminating $\alpha$ between Equations \ref{equ:viscoustimescale} and \ref{equ:thermaltimescale} gives a direct link between the thermal and viscous times,
\begin{equation}\label{equ:viscous_over_thermal}
    \frac{t_\nu}{t_{\rm th}}=\frac{1}{h^2}\quad\Longrightarrow\quad
    h=\sqrt{\frac{t_{\rm th}}{t_\nu}},
\end{equation}
which we use below to infer the inner-disk aspect ratio ($h$) from the observed rise and decay timescales.

It must be emphasised that the observable photometric rise is generally longer than the local thermal time because heating fronts have finite widths and finite propagation speeds and because ionization energy must be supplied in the transition region; heating-front simulations therefore show that measured rises correspond to several local thermal times rather than exactly one $t_{\rm th}$ \citep{1994ApJ...427..987B, 1998MNRAS.298.1048H,1999ApJ...520..276M,2001A&A...373..251D}. To conservatively bracket this effect, we adopt
\begin{equation}\label{equ:thermal_to_rise}
    t_{\rm th}\simeq\frac{t_{\rm rise}}{f},\qquad f\in[2,5],
\end{equation}
i.e. the photometric rise is assumed to be a few times the local thermal time; the range $f=2-5$ reflects the spread of behaviors seen in heating-front calculations \citep{1998MNRAS.298.1048H,1999ApJ...520..276M}.

\subsubsection{Numerical estimates using the decay time}\label{subsubsec:TIDS_numericaltimescales}

We use only the observed decay time as the proxy for the viscous time $t_\nu$. For the 2024 event, the measured decay is $t_{\rm decay}\simeq350$ days; for the 2019 event, the available coverage gives limiting decay values in the range $t_{\rm decay}\simeq128-238$ days. With $t_{\rm rise}\simeq17$ days, and $t_{\rm th}=t_{\rm rise}/f$, we infer the disk aspect ratio $h=\sqrt{t_{\rm th}/t_\nu}$.

For the 2024 decay, $t_\nu=t_{\rm decay}=350$ days and $t_{\rm rise}=17$ days, we obtain, taking $f\in[2,5]$,
\begin{equation}
    t_{\rm th}\in[3.4,8.5]\ {\rm days}\quad\Rightarrow\quad
    h=\sqrt{\frac{t_{\rm th}}{t_\nu}}\simeq 0.10\text{--}0.16
\end{equation}
with a central value $h\approx\sqrt{5.67/350}\simeq0.13$ for $f=3$.

Applying the same procedure to the 2019 limiting decays gives larger aspect ratios: for $t_{\rm decay}=128$ days, we find $h\sim0.16-0.26$ (central $\sim0.21$ for $f=3$), while for $t_{\rm decay}=238$ days, we obtain $h\sim0.12-0.19$ (central $\sim0.15$ for $f=3$). Thus, the 2024 event implies a hot-state aspect ratio $h\sim0.10-0.16$, plausibly in the range expected for a thermally ionized, MRI-active inner disk, whereas explaining the shortest 2019 decay requires either a more strongly puffed-up hot disk or a smaller effective radius for the unstable region during that epoch.

\subsubsection{Expected behaviour of DS and TI and quick comparison with the observations}\label{subsubsec:TIDS_compareObs}

The magnetospheric gated-accretion scenario differs from TI in the physical locus and typical times: in DS, the stellar magnetosphere truncates the inner disk at $R_T$, material supplied from larger radii piles up at that boundary and is released episodically when the barrier is overcome \citep{2012MNRAS.420..416D}. The build-up interval is set by the supply to $R_T$, the rise is often rapid (dynamical or magnetospheric times near $R_T$), and the decay is governed by viscous draining of the innermost region. If $R_T$ collapses to very small radii in outburst, DS-driven bursts can have very fast rises (hours to days) and relatively short decays (weeks to months) depending on the hot-state viscosity.

By contrast, TI is a local thermal–viscous runaway whose observable rise is controlled by heating-front propagation and whose decay is the viscous time at the outer edge of the unstable zone \citep{1994ApJ...427..987B}. Consequently, TI naturally produces longer-duration events (months to years) when the unstable region extends to several stellar radii, and the mass accreted corresponds to the mass in the unstable annulus.

Comparing these expectations with the data, the measured rise of $\sim17$ days and the 2024 decay of $\sim350$ days are readily accommodated by TI for plausible hot-state parameters ($\alpha_{\rm hot}\sim0.05-0.15,\ h\sim0.1$); the shorter 2019 decay (128–238 days) can still be reconciled with TI only if the hot-state disk was substantially puffed up or the effective unstable radius was smaller, or if the observed section of the event captured only part of a longer viscous decay. Below, we estimate the $r_{\rm trigger}$ from Equation \ref{equ:viscoustimescale} for plausible values of $\alpha$ in the range of estimated $h$. For comparison, the corotation radius of Gaia24ccy\,B is 4.73$R_\star \equiv 0.017$ au. \\

For 2024 outburst, $t_{\nu} = t_{\rm decay} = 350$ days and, $h = 0.10- 0.16$, 
\begin{itemize}
    \item $\alpha=0.05 \Longrightarrow 3.2R_\star \leq r_{\rm trigger} \leq 5.9R_\star$;
    \item $\alpha=0.10 \Longrightarrow 5.0R_\star \leq r_{\rm trigger} \leq 9.4R_\star$;
    \item $\alpha=0.15 \Longrightarrow 6.6R_\star \leq r_{\rm trigger} \leq 12.3R_\star$;
\end{itemize}

For 2019 outburst, $t_{\nu} = t_{\rm decay} = 128$ days, and $h = 0.16-0.26$,
\begin{itemize}
    \item $\alpha=0.05 \Longrightarrow 3.0 R_\star \leq r_{\rm trigger} \leq 5.8 R_\star$;
    \item $\alpha=0.10 \Longrightarrow 4.0 R_\star \leq r_{\rm trigger} \leq 9.2 R_\star$;
    \item $\alpha=0.15 \Longrightarrow 6.3 R_\star \leq r_{\rm trigger} \leq 12.0 R_\star$;
\end{itemize}
For 2019 outburst, $t_{\nu} = t_{\rm decay} = 238$ days, and $h = 0.12-0.19$,
\begin{itemize}
    \item $\alpha=0.05 \Longrightarrow 3.1 R_\star \leq r_{\rm trigger} \leq 5.75 R_\star$;
    \item $\alpha=0.10 \Longrightarrow 4.9 R_\star \leq r_{\rm trigger} \leq 9.1 R_\star$;
    \item $\alpha=0.15 \Longrightarrow 6.5 R_\star \leq r_{\rm trigger} \leq 12.0 R_\star$;
\end{itemize}

DS remains viable — especially if, during outburst, the truncation radius moves very close to the star and releases a relatively small pile-up mass — but the numerical timescale and aspect-ratio estimates argue that TI is a plausible primary mechanism for the longer 2024 event.

Within the TI framework, adopting $\alpha_{quiescence}=0.04$, the outburst repeating timescale around 0.2$M_\odot$ star is $\approx$5.4 years \citep[ Equation 20 in][]{2024MNRAS.530.1749N}, comparable to $\sim$5 year separation between the two outbursts of Gaia24ccy\,B, should future observations establish this as a recurrence timescale.

\subsubsection{Mass reservoir}\label{subsubsec:TIDS_massreservoir}

The total accreted mass per event is $M_{\rm acc}\simeq10^{-5}M_\odot$ ($\approx 2 \times10^{28}$ g). The integrated mass accretion rate for the observed duration of 2024 outburst is $1.3\times10^{-5}M_\odot$ (from Figure \ref{fig:MassAcc_replicaOutburst}), while the lower limit of the total accreted mass during 2019 outburst is $6.5\times10^{-6}M_\odot$. In the TI picture, this mass must come from the unstable annulus: a simple estimate of the mean surface density required, $\bar\Sigma\simeq M_{\rm acc}/(\pi r^2$), evaluated at $r\simeq6R_\star$ gives $\bar\Sigma\sim5.3\times10^4\ \mathrm{g\ cm^{-2}}$, and for $r_{\rm trigger}=12R_\star$ it is $\bar\Sigma\sim1.3\times10^4\ \mathrm{g\ cm^{-2}}$. While large, this value is not implausible for the very inner regions of a young disk, and TI operating over a realistic annulus width (e.g. $\Delta r\sim0.2-0.4R_\star$) can comfortably supply $\sim10^{-6}-10^{-5}M_\odot$. In the DS picture, the released mass is the pile-up at $R_T$; DS naturally accommodates small-to-moderate masses per burst (Earth-mass scale to super-Earth scale), and therefore the measured $M_{\rm acc}$ is fully consistent with a single DS release. Thus, mass-budget arguments do not uniquely distinguish between TI and DS.

However, the recurrence time provides an additional constraint. The two observed outbursts are separated by $\sim$5 years, where the quiescent accretion rate onto the star is orders of magnitude lower than the burst accretion rate. If the total accreted mass ($M_{acc}$), piled up at $r_{\rm trigger}$, was supplied steadily from the outer disk, the accretion rate beyond $r_{\rm trigger}$ would be $\simeq 2 \times 10^{-6}M_\odot{\rm yr^{-1}}$. Considering $M_{\rm critical}$ as the threshold of pile-up mass, $\dot{M}_{r>r_{\rm trigger}} = 1.5 \times 10^{-6}M_\odot{\rm yr^{-1}}$. This exceeds the peak accretion rate onto the star during the outburst. Thus, the accumulated mass would always exceed $M_{\rm critical}$, leading to a sustained or quasi-continuous outburst, which is not observed. Instead, the system undergoes discrete events that rapidly drain the accumulated reservoir over 145–367 days. This suggests that the accretion rate beyond $r_{\rm trigger}$ is non-steady or episodic.

With $\dot{M}_{r>r_{\rm trigger}} = 1.5 \times 10^{-6}M_\odot{\rm yr^{-1}}$, Equation 15 of \citet{2024MNRAS.530.1749N} suggests $r_{\rm trigger} \approx 0.043\ \text{au} \equiv 11.3R_{\star}$, consistent with the estimates in Section \ref{subsubsec:TIDS_compareObs}.

In both the TI and DS pictures, some form of temporary mass storage is required: in TI, mass builds up locally until a thermal runaway is triggered, while in DS the accumulation occurs near the truncation radius until magnetospheric conditions allow rapid release. Thus, while the mass reservoir itself does not discriminate between mechanisms, the combination of discrete recurrence, rapid draining, and low quiescent accretion strongly favors scenarios involving episodic inner-disk mass accumulation rather than a steady inflow.

\subsubsection{Magnetospheric evolution and spectral diagnostics}\label{subsubsec:TIDS_spectraldiagnostic}

The magnetospheric truncation radius scales as $R_T\propto\dot M^{-2/7}$ \citep{2007prpl.conf..479B}. The quiescent and peak accretion rates ($\dot M_{\rm q}=2.34\times10^{-9}M_\odot\rm yr^{-1}$ and $\dot M_{\rm peak}=1.74\times10^{-7}M_\odot\rm yr^{-1}$) yields
\begin{equation}
    \frac{R_{T,\rm outburst\ peak}}{R_{T,\rm q}}=\left(\frac{\dot M_{\rm q}}{\dot M_{\rm outburst \ peak}}\right)^{2/7}\simeq0.29,
\end{equation}
so that a quiescent truncation radius, $R_{T,\rm q}= 4.73R_\star$ (corotation radius) would shrink to $R_{T,\rm out}=1.45R_\star$ during the outburst. Such strong inward motion has two immediate observational consequences. First, material accreting very close to the stellar surface produces a dense permitted-line spectrum and strong veiling; the observed emergence of a rich emission-line forest at outburst peak that fades in quiescence (Figure \ref{fig:HFOSCspectra}) is therefore consistent with substantial magnetospheric compression during outburst. Second, if the bulk of accretion proceeds from these small radii, the decay time is set by the (shorter) viscous time at $R_{T,\rm out}$, and the burst would be correspondingly shorter than a TI event that drains a larger-radius annulus.

\subsubsection{Mid-IR colour evolution and hybrid interpretation}\label{TIDS_hybridinterp}

The observed mid-IR color evolution — becoming redder during the outburst (Section \ref{subsec:blueing_reddening}) — indicates that the disk itself is heated and contributes strongly to the mid-IR flux (see Section \ref{subsec:viscousmodel}). This behavior favors a disk-origin heating (i.e., TI) as the primary driver of the luminosity increase rather than a strictly magnetospheric (stellar-surface) burst. 

Taken together with the spectral evidence for magnetospheric compression, the data favor a hybrid interpretation in which a thermally driven enhancement of disk accretion (TI) provides the primary energy and mass release while magnetospheric effects (DS-like gated-accretion or partial truncation collapse) operate during the peak and shape the spectral signatures. We therefore present TI as the more plausible primary mechanism for the 2024 event, with DS effects providing a complementary role; the shorter 2019 event may require different hot-state parameters or a more dominant DS contribution.

\subsubsection{Clump-mediated episodic accretion: a natural route to variable inner-disk mass} \label{subsubsec:ClupmPileUp}

The inferences in Section \ref{subsubsec:TIDS_massreservoir} argue against steady, uniform accretion supplying $M_{\rm critical}$ and instead favour an episodic accumulation of material in the inner disk. A very natural physical channel for such behaviour is the formation and inward migration of dense clumps (or vortices) that act as transient mass reservoirs. Large-scale (magneto)hydrodynamical instabilities — for example Rossby-wave instabilities, zonal flows or vertical-shear instabilities — readily generate long-lived pressure traps and vortices that concentrate gas and solids and can produce planet-scale clumps at radii comparable to $r_{\rm trigger}$ \citep[][and references therein]{1999ApJ...513..805L,2009ApJ...697.1269J, 2020MNRAS.499.1841M}. Such structures offer a robust way to collect similar amounts of mass repeatedly: vortices have a quasi-equilibrium geometry that tends to set the trapped mass and density contrast, and migration or tidal stripping of these clumps can then deliver a comparable mass to the inner disk on each episode.

Finally, clump infall also provides a natural explanation for the shape of the 2024 event: the first phase (till day 118 in Figure \ref{fig:MassAcc_replicaOutburst}) can be powered by mechanism piling up the matter similar to that of 2019 outburst, while the prolonged tail (after day 205 in Figure \ref{fig:MassAcc_replicaOutburst}) could be sustained by episodic delivery of additional clumps or fragments. We therefore consider clump-mediated, episodic mass delivery to be a very plausible — and physically motivated — mechanism for producing the observed, repeatable outbursts.

\subsection{MIR reddening during the outburst: Consequence of Viscous/Irradiated disk}\label{subsec:viscousmodel}

It has long been observed that the YSOs become optically blue when brightness enhances \citep{2011A&A...527A.133K,2012ApJ...749..188L}, however, MIR color has shown myriad features during the brightening events \citep{2021ApJ...920..132P}. This behavior of MIR color has been explained by an increase in circumstellar extinction and a change in the intrinsic color of the disk \citep{2022ApJ...936..152L}.  

An enhanced extinction would suppress the shorter wavelength fluxes more than those at longer wavelengths. This might explain the reddening of YSOs at MIR wavelengths during the outburst; however, the contemporary blueing in the optical wavelengths is in sharp contrast to the extinction. 

\citet{2022ApJ...936..152L} modeled the optical and MIR color evolution of a star-disk system during an increase in the accretion rate. The authors modeled the star and disk using BT-Settl spectra of corresponding temperatures, and a hotspot with a blackbody of 8000 K. The disk temperature was estimated from a combination of irradiation and viscous heating of the disk. The MIR color evolution from this model showed that the color initially becomes redder with increasing mass-accretion rate and then turns blue. Since the time resolution of NEOWISE is insufficient to trace the full locus \citep[from Figure 16 in][]{2022ApJ...936..152L}, the effective color-change observed during the outburst would depend on two factors: initial accretion rate and accretion rate change during the outburst.  

We modeled the emission from Gaia24ccy system across the outburst. Instead of a sophisticated model from \citet{2022ApJ...936..152L}, we modeled the system with a set of blackbodies and calculated the optical and MIR light curves from quiescence to outburst (see Figure \ref{fig:Model_colormag}). The model is described in detail in Appendix \ref{app:viscousdisk}.

\begin{figure}
    \centering
    \includegraphics[width=1\linewidth]{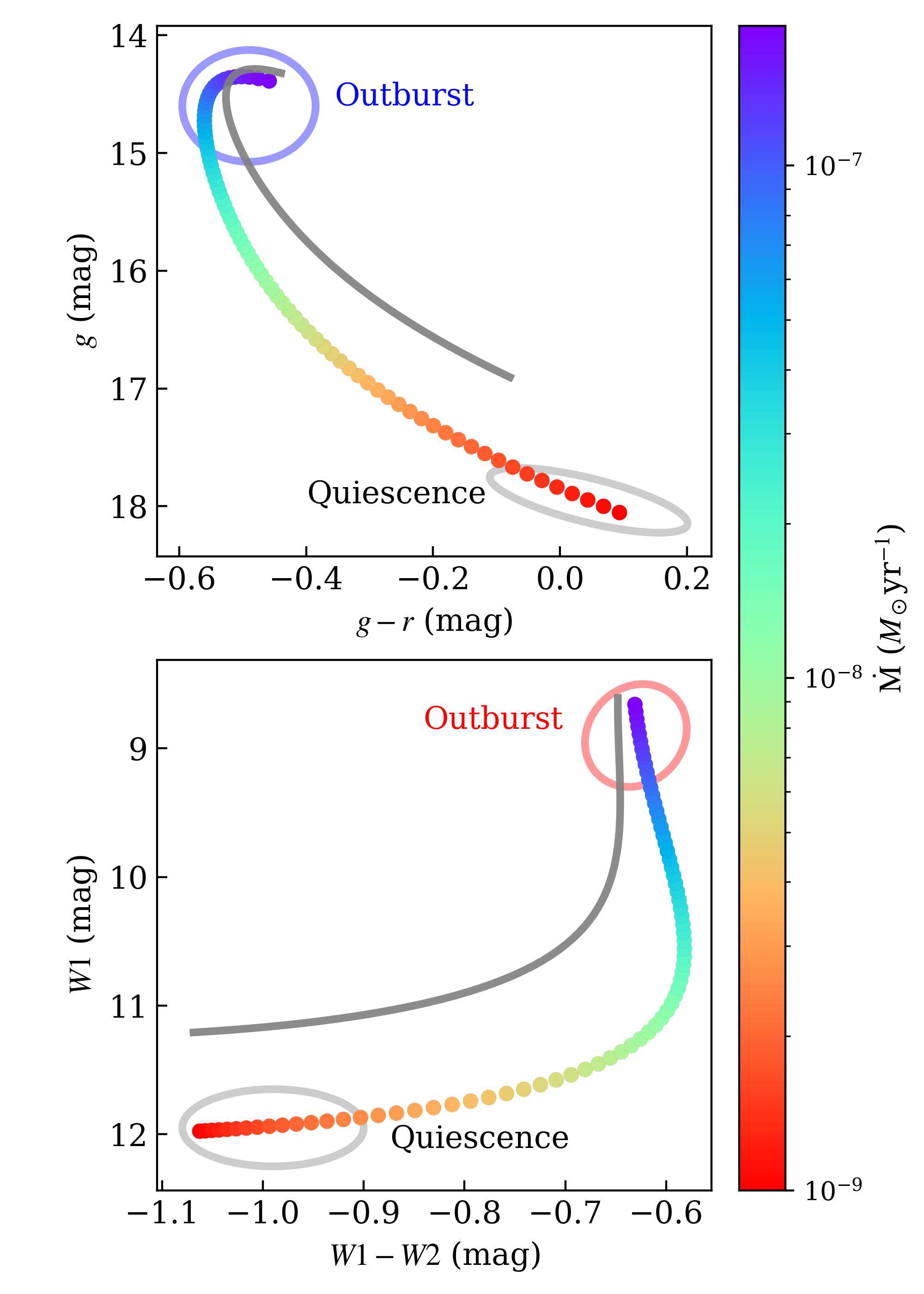}
    \includegraphics[width=1\linewidth]{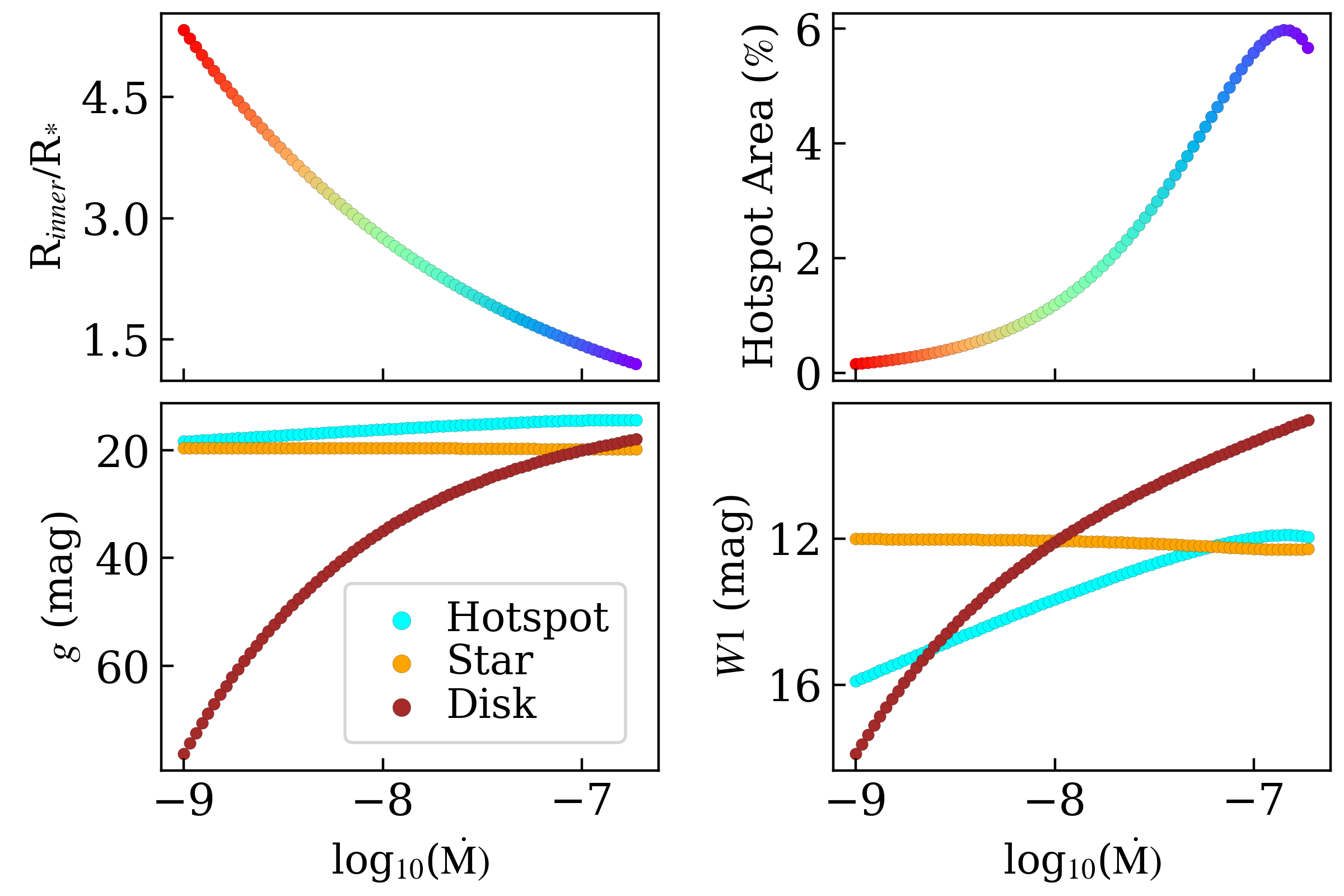}
    \caption{Results from a viscous disk model by varying the mass-accretion rate from $1 \times 10^{-9}$ - 1.9 $\times 10^{-7} M_\odot\,\mathrm{yr}^{-1}$. The top two plots show the optical (\textit{upper}) and MIR (\textit{middle}) color-magnitude evolution for Gaia24ccy\,B (colored dots) and Gaia24ccy (grey curve). Fluxes are converted to magnitude with arbitrary zero points to mimic observations. The colorbar maps the accretion rate to the color of the dots. The \textit{third row} shows the evolution of the inner disk radius (\textit{left}) in terms of stellar radius and fractional area of the stellar surface covered by the hotspot (\textit{right}). \textit{Third row} highlights the contribution of the star, hotspot, and disk in the optical (\textit{left}) and MIR (\textit{right}) fluxes of Gaia24ccy\,B.}
    \label{fig:Model_colormag}
\end{figure}

The results of the model are shown in Figure \ref{fig:Model_colormag}. The first row shows optical ($g-r$) and MIR ($W1-W2$) color-magnitude evolution as mass-accretion rate increases from 10$^{-9} M_\odot\,\mathrm{yr}^{-1}$ during the quiescence to 1.9 $\times$ 10$^{-7} M_\odot\,\mathrm{yr}^{-1}$, during the outburst peak. Colored dots represent Gaia24ccy\,B, while the grey curve highlights the evolution of the combined flux, Gaia24ccy\,A + Gaia24ccy\,B. Initially, Gaia24ccy\,B, and Gaia24ccy become optically bluer as the mass-accretion rate increases. However, both take a sharp turn to become redder at around $\dot{M}_{\rm acc} = 5.7\ \text{and}\ 7 \times 10^{-8} M_\odot\,\mathrm{yr}^{-1}$, respectively. The MIR color evolution shows an intriguing profile: the system initially becomes redder as mass accretion rate increases, and then takes a sharp turn to become bluer after an accretion rate of $\dot{M}_{\rm acc}=2\ \text{and}\ 4.6\ \times 10^{-8} M_\odot\,\mathrm{yr}^{-1}$ for Gaia24ccy\,B and Gaia24ccy, respectively. This suggests that if the objects were accreting at a rate higher by an order of magnitude during the quiescence, the outburst would have shown the objects becoming bluer in the MIR color. The reddening in the optical color might not be observed, given the intrinsic variability in the observed colors. This might provide a qualitative explanation for the findings of \citet{2025MNRAS.tmp..107M} that a larger fraction of Class II objects show MIR reddening during an outburst, while most of the Class I objects demonstrate blueing in the MIR color.

Figure \ref{fig:Model_colormag} shows the optical color changed by $\Delta (g-r) = -0.55$ mag for Gaia24ccy\,B and by $\Delta (g-r) = -0.36$ mag for Gaia24ccy, which is different from the observed $\Delta (g-r) = -0.85$ mag (Section \ref{subsec:blueing_reddening}). The difference could be owed to a large spread in the observed color change (Figure \ref{fig:color-mag}). The modeled MIR color change during the outburst is $\Delta(W1-W2) = 0.43$ mag for Gaia24ccy\,B, and 0.42 mag for Gaia24ccy. This is in good agreement with the observed value of $\Delta(W1-W2) = 0.33$ mag. This suggests that the MIR emitting regions, primarily the two disks, have similar physical conditions. However, the optical flux-emitting regions, hotspots, and photospheres could have differences between the two objects. Furthermore, we must be cautious not to over-interpret this simplistic toy model of the Gaia24ccy system.

The middle row of the Figure \ref{fig:Model_colormag} shows an evolution of the inner disk radius in units of stellar radius, and the fractional area of the stellar surface covered by the hotspot. The bottom row shows the flux contribution in the optical ($g$) and MIR ($W1$) bands by each of the components of Gaia24ccy\,B: Star, Hotspot, and Disk. Hotspot dominates in the optical wavelength; however, in MIR, the disk overshines the central star at an accretion rate higher than $\dot{M}_{\rm acc}= 10^{-8} M_\odot\,\mathrm{yr}^{-1}$, similar to the accretion rate when Gaia24ccy\,B turned bluer in MIR colors. This indicates that the reddening in MIR colors is a consequence of competitive emission between the disk and the stellar photosphere. The disk overshines the hotspot in MIR, for accretion rate higher than 2.5 $\times 10^{-9} M_\odot\,\mathrm{yr}^{-1}$.

\section{Conclusion}\label{sec:conclusion}

We performed a detailed photometric study of an intriguing system, Gaia24ccy, with two lock-headed $\Delta g = 3.8$ mag outbursts at an interval of $\sim$5 years, in 2019 and 2024. The system also experienced a rising MIR and $\Delta V =1$ mag bump in 2014, probably indicating a missed outburst or a regular variability. The system comprises two fast-rotating objects, Gaia24ccy\,A and Gaia24ccy\,B, with a rotation period of $P_{A}$=1.1419 and $P_{B}$=1.7898 days, respectively. The prominent appearance of Gaia24ccy\,B's period just before the 2019 and 2024 outbursts, together with the rise of mean and peak-to-peak flux levels, suggested that the outbursting object was Gaia24ccy\,B. This established Gaia24ccy\,B as a YSO. Gaia24ccy\,A appears as a quasi-periodic dipper, suggesting it to be a highly inclined YSO. Gaia24ccy\,B underwent an accretion rate enhancement by a factor of $\sim$67\ and $\sim$61 during the 2019 and 2024 outbursts, respectively. The two outbursts' profiles showed a remarkable repeatability: both rose at the same rate for the initial 15 days, dipped, and stayed at $g \sim$13 mag until day 80. This was followed by a sharp dip at day 100. The two outbursts exhibited multiple sub-bursts on the time scale of 5-25 days. These sub-bursts are likely regulated by the disk processes. The 2019 outburst lasted 145-255 days, while the 2024 outburst extinguished after 367 days.

\begin{deluxetable}{cc} \tablecaption{Summary table of the outbursts of Gaia24ccy\,B\label{table:conclusion}}
\tablehead{Parameter & Values}
\startdata
Time of outburst & May 2019 and June 2024 \\
Outburst duration & 145-255 days and 367 days \\
Outburst Magnitude & $\Delta g=3.8$ mag \\
Outburst Peak Acc. Rate & 1.7 $\times 10^{-7} M_\odot\,\mathrm{yr}^{-1}$ \\
Quiescence Acc. Rate & 2.3 $\times 10^{-9} M_\odot\,\mathrm{yr}^{-1}$ \\
Mass accumulated (2019) & 6.5 $\times 10^{-6}M_\odot$ \\
Mass accumulated (2024$^{a}$) & 7.5 $\times 10^{-6}M_\odot$ \\
Mass accumulated (2024$^{b}$) & 1.3 $\times 10^{-5}M_\odot$ \\
Disk Radius Involved & $r_{\rm trigger}\sim$ 0.019 $-$ 0.047 au \\
Critical Mass for Outburst & 7.6 $\times 10^{-6}M_\odot$ \\
Acc. Rate$^{c}$ at $r>r_{\rm trigger}$ & 1.5 $\times 10^{-6} M_\odot\,\mathrm{yr}^{-1}$ \\
\hline
Possible Cause of Outburst &  Pile up at $r_{\rm trigger}$ by inhomo-\\
 & -geneous accretion at $r>r_{\rm trigger}$ \\
\enddata
\tablecomments{$^{a}$ mass accumulated in the duration, same as that of 2019. $^{b}$ mass accumulated in the entire 2024 outburst duration. $^{c}$ assuming uniform accretion rate at $r>r_{\rm trigger}$.}
\end{deluxetable}

\begin{figure*}
    \centering
    \includegraphics[width=1\linewidth]{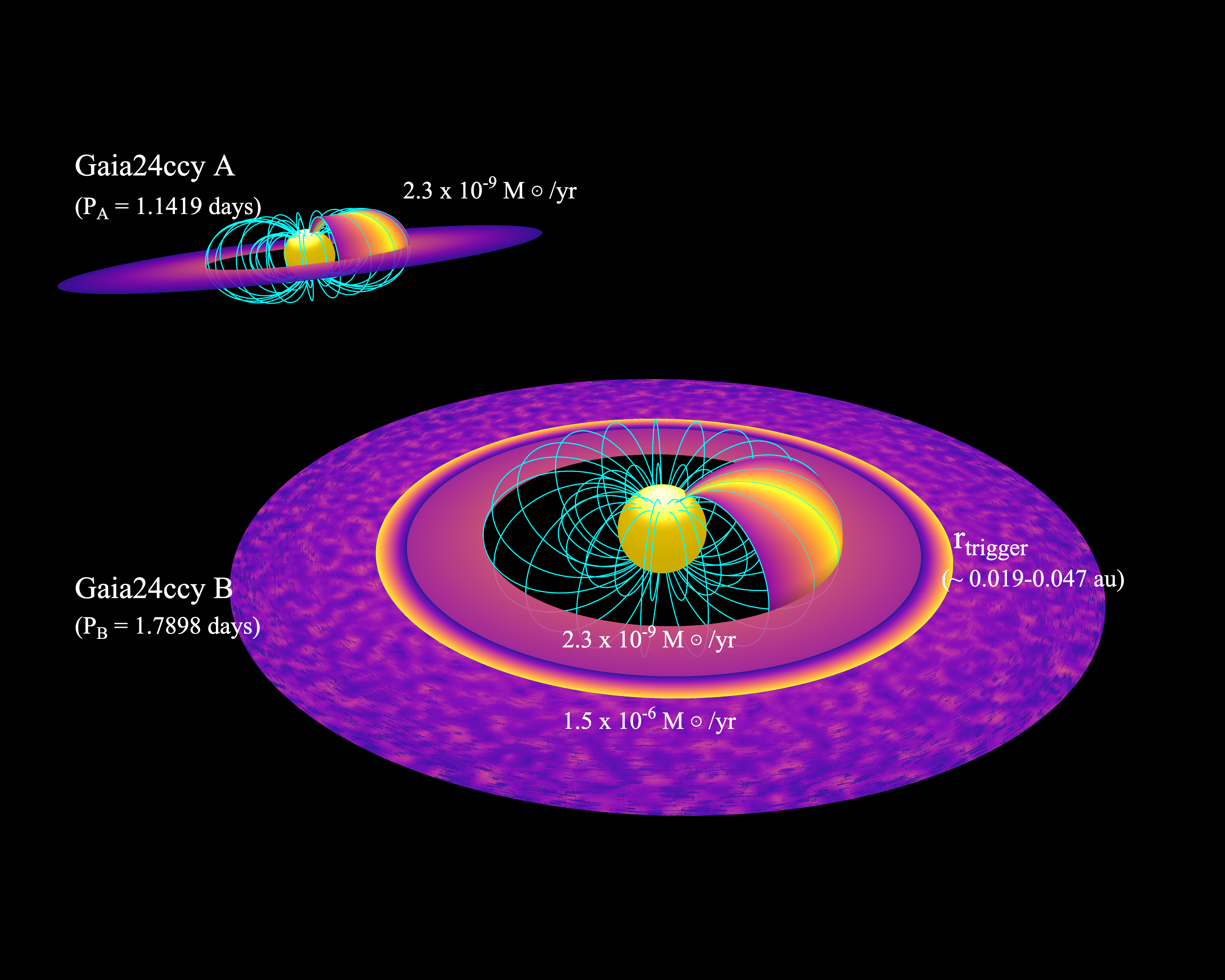}
    \caption{Schematic picture of Gaia24ccy system with two closeby YSOs (Section \ref{subsec:YSOsnature}): Gaia24ccy\,A and Gaia24ccy\,B (Appendix \ref{app:K2_period}). Gaia24ccy\,A has a rotation period of $P_{A}$=1.1419 days and is inclined nearly edge-on, 86$\degree$ (Appendix \ref{app:dipper}). It is accreting at a rate of 2.3 $\times10^{-9}$ $M_\odot\,\mathrm{yr}^{-1}$ (Sections \ref{subsec:stellarparams} and \ref{subsec:massaccretionrate}). Gaia24ccy\,B, shown at the front, rotates with a period of $P_{B}$=1.7898 days. Its inclination angle is unknown; however, we adopted an average value of 57$\degree$. Its accretion rates at the outer and inner disks could be different (Sections \ref{subsec:massaccretionrate} and  \ref{subsubsec:TIDS_massreservoir}). Gaia24ccy\,B accretes at a quiescent rate of 2.3 $\times10^{-9}$ $M_\odot\,\mathrm{yr}^{-1}$ from the inner disk. However, the uniform accretion rate farther away in the disk would be 2 $\times10^{-6}$ $M_\odot\,\mathrm{yr}^{-1}$, causing a pile up of the matter at a radius $r_{\rm trigger} \sim 0.019-0.047$ au (Sections \ref{subsubsec:Reset_mechanism} and \ref{subsubsec:TIDS_compareObs}). However, accretion at $r > r_{\rm trigger}$ could be inhomogeneous to contain the finite duration of the outbursts (Section \ref{subsubsec:ClupmPileUp}), where inhomogeneity could be in the form of clumps, as illustrated in the figure. The piled-up matter would reach a critical mass ($M_{\rm critical} = 7.6 \times 10^{-6}\, M_\odot$) in $\sim$5 years, causing an accretion outburst (Section \ref{subsubsec:Reset_mechanism}).
    }
    \label{fig:schematic_Gaia24ccy}
\end{figure*}

The similarity of the outburst profiles during rise and most of the decay, along with a similar amount of mass accretion, constrained the trigger mechanism to TI or DS. Comparing the thermal and viscous timescales to the rise and decay timescales of the two outbursts, we estimated the disk scale height ratio ($h$), which appears plausible ($h = 0.10 - 0.16$) for a hot and puffed-up disk during the 2024 outburst. However, reconciling it with the 2019 event would require either a highly puffed disk ($h = 0.16-26$) or a smaller trigger radius, $r_{\rm trigger}$. For a plausible hot disk value of $\alpha \in (0.10 - 0.15)$, we found the unstable region of the disk to be compact, $5.0R_\star (\equiv 0.019\ {\rm au}) \leq r_{\rm trigger} \leq 12.3R_\star (\equiv 0.047\ {\rm au})$. The mean surface density of the disk estimated from total accreted mass does not preferentially distinguish between TI and DS. The timescales of the events, emergence of emission spectra, and reddening of MIR color during outburst suggest the trigger mechanism for the 2024 outburst to be disk-driven, plausibly TI aided by DS during the peak of the outburst. However, DS appears more dominant mechanism for the 2019 outburst.

The mass-reservoir provided an estimation of steady mass inflow rate beyond the trigger radius, $\dot{M}_{r > r_{\rm trigger}} = 2 \times 10^{-6}$ $M_\odot\,\mathrm{yr}^{-1}$. $\dot{M}_{r > r_{\rm trigger}}$ exceeds the peak mass accretion rate onto the star during the outbursts, suggesting that the mass inflow beyond $r_{\rm trigger}$ is inhomogeneous.

Gaia24ccy system showed a reddening in the MIR color during the rise of the 2019 outburst, while turning bluer in the optical color. We modeled the system with a set of blackbodies: star + hotspot + disk. The spectral energy distribution reproduced the observations when the model's accretion rate was increased from the observed quiescence state value to that of the outburst state. The model demonstrated that the observed MIR reddening is driven by emission from the inner-disk dominating over the central star contribution. This further provided a qualitative explanation for the findings that the MIR color of most of the Class I objects becomes blue during the outburst, while that of Class II objects mainly turns redder. 

We suggest continuous monitoring of Gaia24ccy. Confirmation of the next outburst in 2029 would be critical to the 5-year periodicity of the outbursts. Future spatially resolved spectroscopy could help better characterize both stellar objects. Spatially resolved photometry would be essential to understand the binarity of the system, for its position angle might have evolved. High-resolution ALMA observations would be essential to constrain whether the two disks are circumstellar or circumbinary, and whether they are misaligned. 

We conclude by tabulating the findings in Table \ref{table:conclusion} and by presenting a schematic of Gaia24ccy system in Figure \ref{fig:schematic_Gaia24ccy} with two disk-bearing YSOs, one at the front undergoing outburst and the second at the back staying dormant.

\section{Acknowledgements}
We are thankful to the reviewer for the detailed and constructive feedback that helped us improve the clarity and quality of the paper. We thank Prof. Jerome Bouvier for the discussion on the dippers. K.S., J.P.N., and D.K.O. acknowledge the support of the Department of Atomic Energy, Government of India, under Project Identification No. RTI 4002. K.S. acknowledges the financial assistance from the Science \& Engineering Research Board (SERB), Government of India, for this work, supported by grant no. ITS/2024/004263. K.S. acknowledges the Infosys-TIFR Leading Edge Travel Grant. This research was supported in part by a generous donation (from the Murty Trust) aimed at enabling advances in astrophysics through the use of machine learning. The Murty Trust, an initiative of the Murty Foundation, is a not-for-profit organization dedicated to the preservation and celebration of culture, science, and knowledge of systems born out of India. The Murty Trust is headed by Mrs. Sudha Murty and Mr. Rohan Murty. This work is supported by the China-Chile Joint Research Fund (CCJRF No.2301) and the Chinese Academy of Sciences South America Center for Astronomy (CASSACA) Key Research Project E52H540301. ZG, RK, JB, and JO are funded by ANID, Millennium Science Initiative, AIM23-001. ZG and CM are funded by the project ALMA-ANID 31240014. J.B. and R.K. thank the support from FONDECYT Regular project No. 1240249. This work is supported by the Fundamental Fund of Thailand Science Research and Innovation (TSRI) through the National Astronomical Research Institute of Thailand (Public Organization) (FFB680072/0269). VE acknowledges support from the Ministry of Science and Higher Education of the Russian Federation (State assignment in the field of scientific activity 2023, GZ0110/23-10-IF). We thank the staff at IAO, Hanle, and CREST, Hosakote,
operated by the Indian Institute of Astrophysics, Bengaluru (India), for their support in facilitating these observations. This paper includes data gathered from the Three-hundred MilliMeter Telescope (TMMT) and the Las Campanas Remote Observatory (LCRO) located at Las Campanas Observatory, Chile. We thank Barry Madore for initiating the TMMT project. This research is based on observations made with the Thai Robotic Telescope under program ID TRTC12A\_013, which is operated by the National Astronomical Research Institute of Thailand (Public Organization). This work makes use of observations from the Las Cumbres Observatory global telescope network. \\

\appendix 

\section{Two photometric periods in K2 light curve}\label{app:K2_period}

\citet{2018AJ....155..196R} reported a stellar rotation period of 1.1427 days for Gaia24ccy by analyzing the K2 light curve, obtained during 2014 August 24 to 2014 November 11. This observation duration is immediately after the `2014 burst/outburst?' bump in Figure \ref{fig:lightcurve}.

\begin{figure}
    \centering
    \includegraphics[width=1\linewidth]{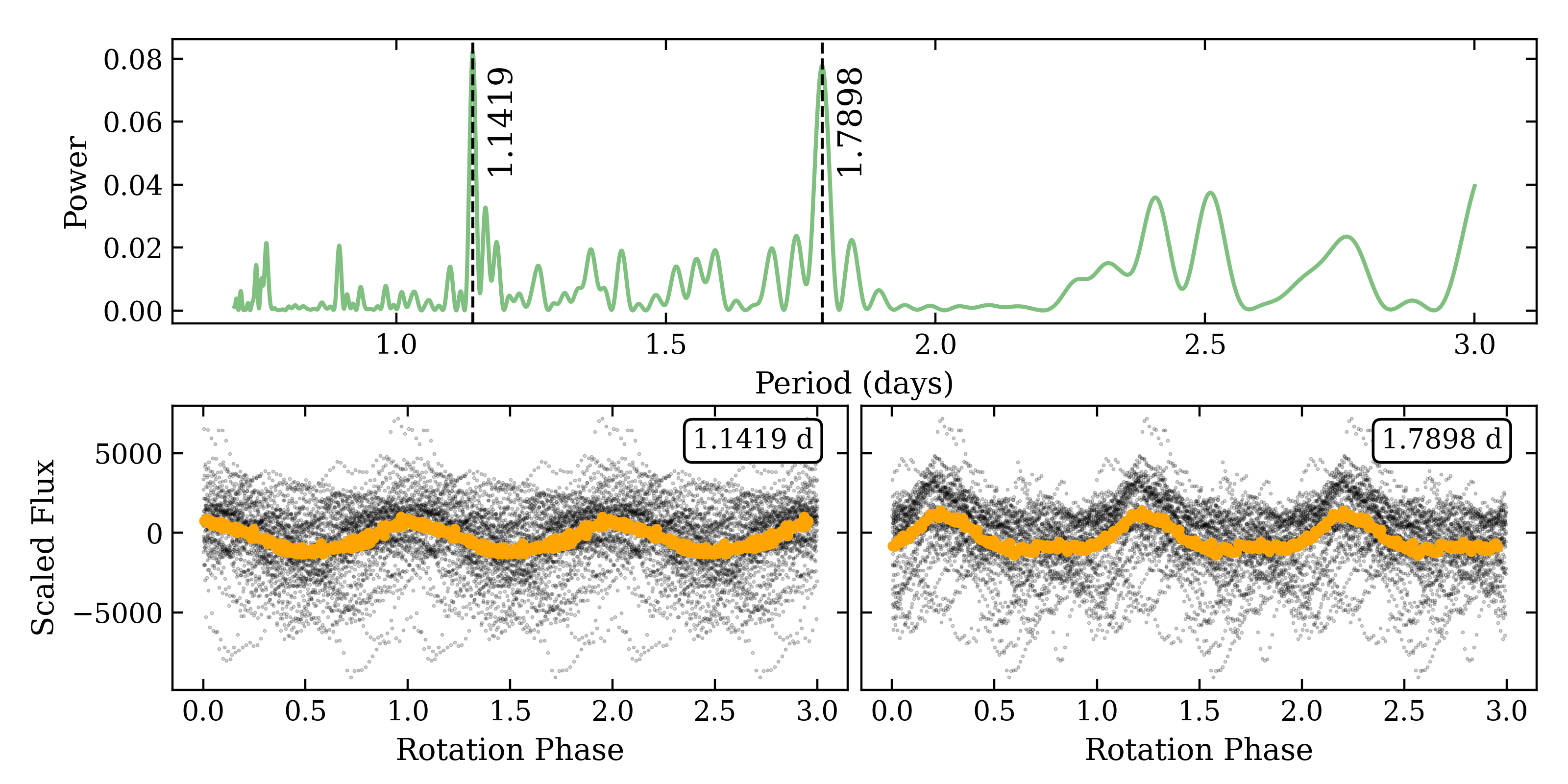}
    \caption{\textit{Upper} panel shows the LombScargle power spectrum of the K2 light curve. Two peaks at $P_{A}$=1.1419 days and $P_{B}$=1.7898 days are marked with black dashed vertical lines. Grey dots in the $lower\ panel$ are the phase-folded K2 light curve at both periods. Orange points are the running median of the grey dots.}
    \label{fig:K2_2fold}
\end{figure}

We also performed the Lomb-Scargle periodogram analysis over the same K2 light curve. The power spectrum showed two prominent peaks within 1-3 days, at $P_{A}=$1.1419 and $P_{B}=$1.7898 days, with no significant peaks up to 15 days. The false alarm probabilities of both periods are essentially zero. Both periods fold the light curve well (Figure \ref{fig:K2_2fold}). The phase-folded light curve at $P_{A}$ is nearly sinusoidal, while it appears sinusoidal at $P_{B}$ for $\sim$50\% rotation cycle and then becomes flat (Figure \ref{fig:K2_2fold}). 

Indeed, the $K'$-band adaptive optics (AO) image of the system revealed two photometrically similar ($\Delta K=0.26$ mag) point-like objects at a close proximity \citep{2019ApJ...878...45B}. The projected angular distance between the objects is 138.4 $\pm$ 1.8 milliarcsec (mas), corresponding to 19.7 au. We therefore associate the two periods with these two components. Hereafter, we refer to the first object as Gaia24ccy\,A, with $P_{A}$ rotation period, and the second object as Gaia24ccy\,B, with $P_{B}$ rotation period. The combined system, Gaia24ccy\,A + Gaia24ccy\,B, will be referred to as Gaia24ccy. Figure \ref{fig:KeckAO} shows the AO image from \citet{2019ApJ...878...45B}. In the absence of spatially resolved time series photometry, we cannot determine whether the upper object in Figure \ref{fig:KeckAO} corresponds to Gaia24ccy\,A/B, or whether it is the lower object.

\section{Appearance of $P_{B}$ just before the outbursts: Identifying Gaia24ccy\,B as the outbursting component}\label{app:asasn_period}

\begin{figure*}
    \centering
    \includegraphics[width=1\linewidth]{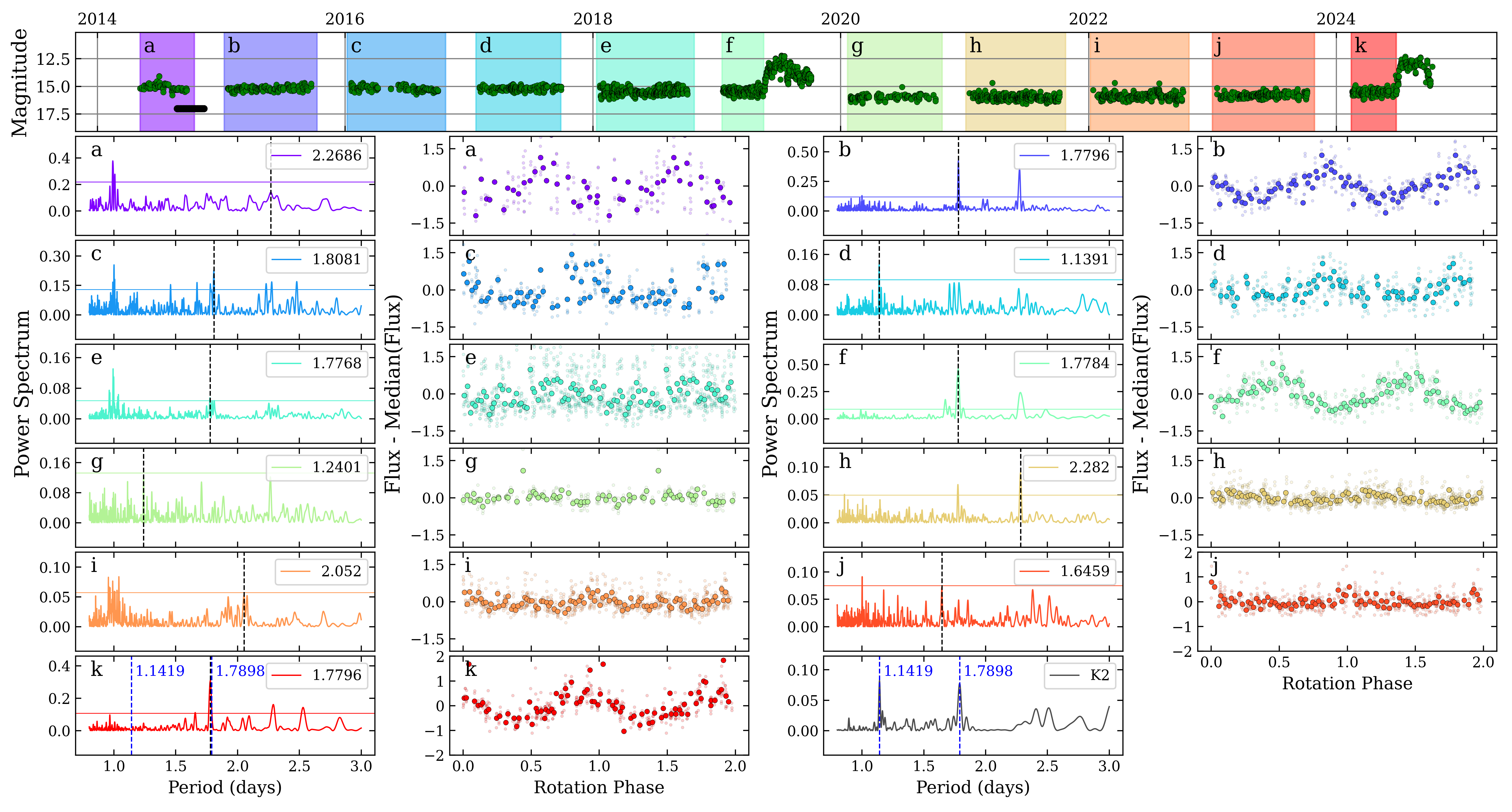}
    \caption{Lomb-Scargle periodogram analysis of ASAS-SN $g$ and $V$-band light curves of Gaia24ccy is shown. The \textit{top panel} presents the combined ASAS-SN \textit{V} and \textit{g}-band light curves shown by green dots. The shaded time windows are analyzed for periodicities separately. \textit{Bottom panels}: Sets of adjacent panels show the Lomb-Scargle power spectrum and corresponding phase-folded light curve over two rotations. An English alphabet in the shaded regions of the light curve and power spectrum marks the correspondence. A one percent false alarm probability level is indicated by a horizontal line in each of the power spectra. A vertical dashed black line shown in the power spectrum marks the location of the recovered period, which are also written at the top right corner of the respective power spectra. The power spectra of the K2 light curve from Figure \ref{fig:K2_2fold} are shown in the lower right panel, with $P_{A}$ and $P_{B}$ marked for reference. The same two periods are also highlighted with vertical dashed blue lines in the lowest left panel. The phase-folded light curves are shown with faint dots, and their running mean are the brighter dots.}
    \label{fig:ASASN_periodogram}
\end{figure*}

ASAS-SN observed Gaia24ccy across several observation windows\footnote{The term `window' indicates a full observation span before the star hides behind the sun.}, from 2014 May 7 to the quiescent phase following the 2024 outburst. We analyzed the Lomb-Scargle periodogram for each window separately (see Figure \ref{fig:ASASN_periodogram}). The power spectra along with 1\% false alarm probability (FAP) levels and phase-folded light curves for each window are shown in Figure \ref{fig:ASASN_periodogram}. ASAS-SN light curves, though sparse for such short-period objects, picked up the rotation periods found in the K2 light curve. Periods within 0.05 days of 1 day were excluded to mitigate aliasing from the daily sampling cadence. Periodic signals from phase-folded light curves are apparent on a few windows, marked \textbf{a} (P=2.268 days), \textbf{b} (P=1.7796 days), \textbf{f} (1.7784 days), \textbf{k} (1.7796 days), and probably \textbf{h} (P = 2.282 days). 
A power spectrum peak around $P_{B}$ appeared in almost every window; however, they are significantly above 1\% FAP only in windows \textbf{b, f,} and \textbf{k}. The light curves folded at $P_{B}$ also did not appear periodic in many windows, except for those in \textbf{b, f,} and \textbf{k}. Windows \textbf{f} and \textbf{k}, with periods similar to that of Gaia24ccy\,B, were observed just before the `outburst 2019' and `outburst 2024' (see Figures \ref{fig:lightcurve} and \ref{fig:ASASN_periodogram}). Window \textbf{b}, with period similar to $P_{B}$, was observed immediately after `2014 outburst?'; although, no such periodic signal appeared in the window \textbf{g}, observed immediately after the 2019 outburst. The maximum difference between the periods in windows \textbf{b}, \textbf{f}, \textbf{k} and $P_{B}$ is only 2$\degree$ of the stellar rotation, suggesting them to be the rotation signal of Gaia24ccy\,B. Thus, the preferential appearance of $\approx P_{B}$ period just before the 2019 and 2024 outbursts suggests that Gaia24ccy\,B is the outbursting object. Spatially resolved observations would be required to confirm this identification unambiguously. 

The appearance of $P_{B}$ in windows \textbf{f} and \textbf{k} could indicate an initial accretion-driven brightening of the star even before the apparent onset of the outbursts. We indeed found that both the peak-to-peak intrinsic variability and mean flux level increased in the windows \textbf{f} and \textbf{k} compared to the previous windows. The median flux levels (and peak-to-peak flux level) in the windows \textbf{f} and \textbf{k} were 2.45 (1.96) mJy and 2.24 (1.79) mJy, respectively, while those in preceding windows, \textbf{e} and \textbf{j}, were 2.13 (1.23) mJy and 1.69 (1.25) mJy, respectively. A slight increase in the accretion rate before the actual onset of the outburst would enhance the hotspot contribution to the photometric flux. This can increase the peak-to-peak photometric value, along with raising the mean photometric level. Thus, the above observations might indicate that the accretion rate began to increase even before the actual onset of the outbursts. This could potentially help trace, anticipate, and prepare for the upcoming outbursts. From hereonwards, we consider Gaia24ccy\,B as the outbursting object, while Gaia24ccy\,A remains dormant.

The power spectra also picked periods $\sim$2.28 days on almost every window, but the peaks are above 1\% FAP only when $\approx P_{B}$ peaks are above 1\% FAP. Indeed, the power spectra of a sine function with period $P_{B}$ sampled onto the ASAS-SN epochs of Figure \ref{fig:ASASN_periodogram} produce two peaks: $P_{B}$ and $\sim 2.28$ days, indicating that the latter peaks are an alias of $P_{B}$.

Interestingly, periods similar to $P_{A}$ do not appear in any windows, except \textbf{d} (P = 1.139 days), and even there the phase-folded light curve does not show a clear periodic signal. To understand the possible reasons for the absence of $P_{A}$ from Figure \ref{fig:ASASN_periodogram}, we generated synthetic light curves by adding two sine curves with periods $P_{A}$ and $P_{B}$, while varying their amplitude, SNR, and relative phases. We found that the power spectra peak at $P_{A}$ disappear when the peak-to-peak amplitude of Gaia24ccy\,A is $\leq 40\%$ to that of Gaia24ccy\,B. This could be indicative of different accretion rates on both the objects or different inclination angles, both resulting in different hotspot luminosity of Gaia24ccy\,A and Gaia24ccy\,B, producing different peak-to-peak amplitudes in optical light curves. We assumed similar quiescent accretion rates for both objects but adopted different inclination angles (see Sections \ref{subsec:massaccretionrate} and \ref{subsec:viscousmodel}). We used an average inclination angle of 57$^{\circ}$ for Gaia24ccy\,B, and a high inclination angle of 86$^{\circ}$ for Gaia24ccy\,A, following \citet{2017ApJ...851...85B} and Appendix \ref{app:dipper}.

\section{Dips in K2 light curve: Possibility of Gaia24ccy\,A as a disk bearing YSO}\label{app:dipper}

\begin{figure} 
    \centering
    \includegraphics[width=1\linewidth]{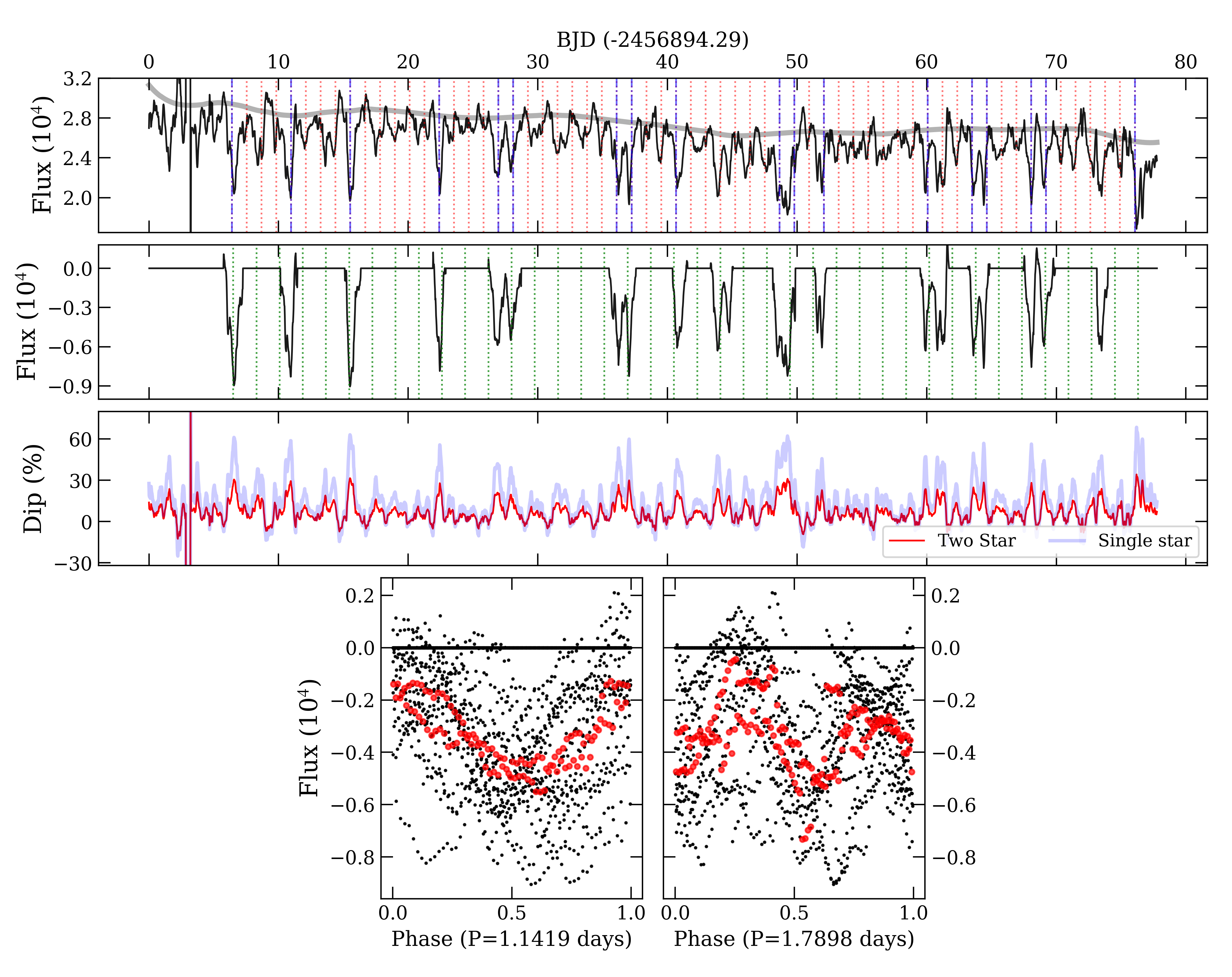}
    \caption{Dips in K2 light curve. \textit{Upper row} shows the K2 light curve with dotted red vertical lines at an interval of $P_{A}$. The red lines overlapping with the dips are highlighted in dashed-dotted blue lines. The smoothed 85th percentile continuum of the light curve is shown with a grey curve. \textit{2$^{nd}$ row} is the continuum-detrended light curve with flux equated to 0 except for the dips. Dashed-dotted green vertical lines are at an interval of $P_{B}$ to highlight that this period misses many dips. \textit{3$^{rd}$ row} shows the fractional percentage of the total flux covered by dips (red curve). Assuming that both stars equally contribute to the total flux, the flux percentage dip of a single star doubles (blue curve). \textit{Bottom row} shows the phase folded light curve from the \textit{2$^{nd}$ row}. \textit{Left} and \textit{right} panels are folded at $P_{A}$ and $P_{B}$ periods, respectively. A running median is shown with red dots.}
    \label{fig:K2_dippers}
\end{figure}

The K2 light curve of Gaia24ccy displays occasional dips as shown in Figure \ref{fig:K2_dippers}. We searched for a periodicity in the dips. The top panel of Figure \ref{fig:K2_dippers} has red dotted vertical lines marked at an interval of $P_{A}$, the rotation period of Gaia24ccy\,A (see Section \ref{app:K2_period}). The first vertical line is phase-shifted to maximize the number of lines overlapping with the dips. The lines that do overlap are highlighted in blue. All the major dips align at this period. However, the minima of the dips do not consistently align, and also, not every rotation cycle exhibits a dip. To see a clear signal of periodicity in the dips, we detrended the light curve using its 85th percentile smoothed continuum. This continuum is shown as a grey curve in the upper panel of Figure \ref{fig:K2_dippers}. The flux values of the detrended light curve are equated to 0 other than for the dips (see 2$^{nd}$ row in Figure \ref{fig:K2_dippers}). We marked the 2$^{nd}$ row with dotted green vertical lines at the intervals of $P_{B}$, the rotation period of Gaia24ccy\,B. However, these green lines did not coincide with several prominent dips. The 3$^{rd}$ row in Figure \ref{fig:K2_dippers} shows the fractional percentage of flux decrement during the dips (red curve). The light curve showed a flux decrease of up to 30\% relative to the baseline. Assuming an equal photometric contribution from both objects, this corresponds to a dip of up to 60\% for a single object. The dips often extended beyond a full rotation cycle of Gaia24ccy\,A.

The dips in light curve of the 2$^{nd}$ row is phase folded at both periods and is shown in the bottom panel of Figure \ref{fig:K2_dippers}. The light curve folded at $P_{A}$ shows a stable dip; however, the dip features got smeared out when folded at $P_{B}$. A running mean of the folded light curves, shown in red dots, enhances the alignment of dips at $P_{A}$. Both the upper and lower panels indicate the association of the dips with $P_{A}$, and so, with Gaia24ccy\,A. The LombScargle periodogram of the dipper light curve (2$^{nd}$ row in Figure \ref{fig:K2_dippers}) in the period range $0.5-5$ days shows a prominent peak at 4.15 days, however the phase-folded light curve is smeared out; while the most prominent peak in period range $0.5-2$ days is at 1.138$\sim P_{A}$, and the phase-folded light curve demonstrate a stable dip similar to the lower left panel of Figure \ref{fig:K2_dippers}. The quasi-periodic nature of the dips could be a result of occultation by non-stable levitation of the disk material at the corotation radius.

To understand the alignment of a few dips with $P_{B}$ (in 2$^{nd}$ row of Figure \ref{fig:K2_dippers}), we constructed dipper light curves with 25 dips placed at random integer multiples of $P_{A}$ with 20\% uncertainty. The full width at half-maxima was kept 120$^\circ \pm30^\circ$ of the rotation of Gaia24ccy\,A. With 50000 such light curves, we found that $\approx$35\% of dips coincide with $P_{B}$. We observed around seven dips aligning with $P_{B}$ out of 20 dips in Figure \ref{fig:K2_dippers}, consistent with the above result. This suggests that under the assumption of association of dips with Gaia24ccy\,A,
the observed alignment of a few dips at $P_{B}$ could be coincidental. 

Thus, for this study, we consider the scenario of Gaia24ccy\,A having a highly inclined disk, where disk extinction produces quasi-periodic photometric dips.

\section{Viscous disk model of Gaia24ccy system}\label{app:viscousdisk}

We modeled each component of the Gaia24ccy system with a set of blackbodies: a central star with a blackbody of photospheric temperature, a hotspot with a blackbody of 9000 K, and the circumstellar disk with blackbodies of viscously heated annuli. We defined the inner disk radius ($R_{in}$) as the point at which the disk ram pressure balances the magnetic pressure. $B_\star$=1 kG showed that the inner disk reached the stellar surface at an accretion rate less than that during the outburst peak. So, we used another reasonable estimate of $B_\star$ by equating the corotation radius and truncation radius during the quiescence. This provided $B_\star\approx1.5$ kG. We also set, $R_{in} = max(R_\star, R_{in})$. We varied the hotspot area to match its luminosity to that expected from the gravitational energy released by the free fall of the accreting material from $R_{in}$ to $R_\star$. We adopted an average inclination angle of 57$^{\degree}$ for Gaia24ccy\,B, and 86$^{\degree}$ for Gaia24ccy\,A \citep{2017ApJ...851...85B}. We varied the mass-accretion rate and collected the fluxes in optical and MIR bands. The fluxes were converted into magnitude scale with arbitrary zero point magnitudes to mimic the observed magnitude ranges in Figure \ref{fig:color-mag}. We combined the luminosities of Gaia24ccy\,A and Gaia24ccy\,B, where Gaia24ccy\,A remains in quiescence while Gaia24ccy\,B undergoes an outburst. The results are shown in Section \ref{subsec:viscousmodel} and Figure \ref{fig:Model_colormag}.

\bibliography{reference}{}
\bibliographystyle{aasjournal}

\end{document}